\documentclass{aa}

\usepackage[T1]{fontenc}
\usepackage{ae,aecompl}
\usepackage[varg]{txfonts}

\usepackage[utf8]{inputenc}

\usepackage{graphicx}
\usepackage{xspace}
\usepackage{bm}

\graphicspath{./fig/} 

\usepackage{ulem,cancel}

\DeclareMathOperator{\im}{Im}	
\DeclareMathOperator{\re}{Re}

\newcommand{\e}{\mathrm{e}}        
\newcommand{\od}{\mathrm{d}}        
\newcommand{\dd}{\mathrm{d}}        
\newcommand\deriv[2]{\frac{\partial#1}{\partial#2}}
\newcommand\oderiv[2]{\frac{\od#1}{\od#2}}

\newcommand{\red}[1]{#1}

\newcommand{\const}{\mathrm{const}}            
  
\newcommand{\h}{_\mathrm{h}}  
\newcommand{\di}{_\mathrm{d}}  
\newcommand{\crit}{_\mathrm{c}}  
\newcommand{\Dlocal}{\mathcal{D}}  
\newcommand{\Roh}{R_{\omega\text{h}}}		
\newcommand{\Rod}{R_{\omega\text{d}}}		 
 \newcommand{\Rah}{R_{\alpha\text{h}}}		
 \newcommand{\Rad}{R_{\alpha\text{d}}}		
 \newcommand{\What}{\widehat{\mathcal{W}}} 

\newcommand{\hz}{\widehat{z}}
\newcommand{\az}{a}

\newcommand\vect[1]{\pmb{#1}}                   
\newcommand{\etat}{\beta}  

\newcommand{\cm}{\,{\rm cm}}    
\newcommand{\km}{\,{\rm km}}    
\newcommand{\m}{\,{\rm m}}      
\newcommand{\pc}{\,{\rm pc}}    
\newcommand{\p}{\,{\rm pc}}     
\newcommand{\kpc}{\,{\rm kpc}}  
\newcommand{\g}{\,{\rm g}}      
\newcommand{\s}{\,{\rm s}}      
\newcommand{\yr}{\,{\rm yr}}    
\newcommand{\kms}{\km\s^{-1}}    
      
\newcommand{\muG}{\,\mu{\rm G}} 
\newcommand{\rad}{\,\mathrm{rad}}

\newcommand{\HI}{H\,\textsc{i}\xspace} 
\renewcommand{\vec}{\bm}

\newcommand{\rcyl}{s} 

\newcommand{\galmag}{\textsc{galmag}\xspace}
\newcommand{\hammurabi}{\textsc{Hammurabi}\xspace}

\usepackage[pdftex,breaklinks=true,pdfborder={0 0 0.05},colorlinks]{hyperref}
\hypersetup{
     pdftitle={A physical model of the GMF},
     pdfauthor={
          A.~Shukurov (orcid.org/0000-0001-6200-4304),
          L.~F.~S.~Rodrigues (orcid.org/0000-0002-3860-0525),
          P.~J.~Bushby,
          J.~Hollins,
          J.~Rachen
     },
      allcolors = {blue}
}

\usepackage{natbib,twoopt}

\begin{document}
\title{A physical approach to modelling large-scale galactic magnetic fields}
\author{Anvar~Shukurov\inst{1} \and
           Luiz~Felippe~S.~Rodrigues\inst{1} \and
           Paul~J.~Bushby\inst{1} \and
           James~Hollins\inst{1} \and
           J{\"o}rg~P.~Rachen\inst{2}
           }

\titlerunning{A physical approach to GMF modelling}
\authorrunning{A.~Shukurov et al.}
\institute{School of Mathematics, Statistics and Physics, Newcastle University,
           Newcastle upon Tyne, NE1 7RU, UK
           \email{
           \href{mailto:anvar.shukurov@ncl.ac.uk}{anvar.shukurov@ncl.ac.uk};
        \href{mailto:luiz.rodrigues@ncl.ac.uk}{luiz.rodrigues@ncl.ac.uk};
        \href{mailto:paul.bushby@ncl.ac.uk}{paul.bushby@ncl.ac.uk};
        \href{mailto:j.hollins@ncl.ac.uk}{j.hollins@ncl.ac.uk};
        }
       \and
           Department of Astrophysics/IMAPP, Radboud University,
           P.O.\ Box 9010, 6500 GL Nijmegen, The Netherlands
           \email{\href{mailto:j.rachen@astro.ru.nl}{j.rachen@astro.ru.nl}}
}

\abstract
{A convenient representation of the structure of the
large-scale galactic magnetic field is required for the interpretation of polarization 
data in the sub-mm and radio ranges, in both the Milky Way and external galaxies.
}
{We develop a simple and flexible approach to construct parametrised models of the 
large-scale magnetic field of the Milky Way and other disc galaxies, 
based on physically justifiable models of magnetic field structure. 
The resulting models are designed to be optimised against available
observational data. 
} 
{Representations for the large-scale magnetic fields in the flared disc and spherical 
halo of a disc galaxy were obtained in the form of series expansions
whose coefficients can be calculated from observable or theoretically
known galactic properties. The functional basis for the expansions is 
derived as eigenfunctions of the mean-field dynamo equation or of the  
vectorial magnetic diffusion equation.
}
{The solutions presented are axially symmetric 
but the approach can be extended straightforwardly to non-axisymmetric cases.
The magnetic fields are solenoidal by construction, can 
be helical, and are parametrised in terms of observable properties of the host
object, such as the rotation curve and the shape of the gaseous disc. 
The magnetic field in the disc can have a prescribed number of field reversals at 
any specified radii. Both the disc and halo magnetic fields can separately have
either dipolar or quadrupolar symmetry. 
The model is implemented as a publicly available software package \galmag which allows, in particular, the computation of the synchrotron emission and Faraday 
rotation produced by the model's magnetic field.
}
{The model can be used in interpretations of observations of magnetic fields in 
the Milky Way and other spiral galaxies, in particular as a prior in Bayesian analyses.  
It can also be used for a simple simulation of a time-dependent magnetic field generated by dynamo action. 
}

\date{Accepted for publication in A\&A}
\maketitle
\section{Introduction}

Recent increased interest in the large-scale magnetic fields of the Milky Way (MW)
and other spiral galaxies is driven by a number of factors. Their role in the
dynamics of the interstellar medium (ISM) has been widely appreciated although not 
completely understood. Their significance for the feedback processes in evolving galaxies 
has also been recognised and is being actively explored \citep[e.g.][]{Fire2}.
Furthermore, the separation of the Galactic foreground, including its magnetic field and associated
emission, from extragalactic contributions is essential to identify the sources of ultra-high 
energy cosmic rays (UHECR) \citep{KoOl11,MoRo18} and for cosmological interpretations of 
sensitive CMB observations \citep[e.g.][]{PlanckXI2018}.

Our understanding of the Galactic magnetic field (GMF) is based on
observations of Faraday rotation of polarised radio emission, synchrotron emission of energetic 
electrons, and polarised emission and absorption by dust \citep[e.g.][]{haverkorn2015}.
Observations of external spiral galaxies have provided rich polarisation data for their 
discs and haloes \citep{Beck15,Wiegert2015,Mao2015}. 
Interpretation of such data in terms of three-dimensional magnetic field structures
requires parametrised models of the magnetic field based on the understanding of their
nature and origin. It can be expected that models that rely less on specific theoretical
models would have a larger number of free parameters whose physical meaning would be less
clear. On the contrary, physically motivated models can be more flexible and lead to
physically transparent interpretations of observations.

A range of heuristic models for the structure of the GMF have been proposed 
\citep[e.g.][]{Sun2008,Jaffe2010,VanEck2011,JF12,JF12a,Ferriere2014,Terral2017}.
Magnetic fields in most of these models are superpositions of various {ad hoc}
parts whose parameters are selected by fitting to a range of observables. The
heuristic nature of the models makes them rather inflexible. Furthermore, their
parameters are not necessarily related to the ISM properties (often lacking a
transparent physical meaning), and some such models fail to satisfy even such
fundamental constraints as $\nabla\cdot\vec{B}=0$ or to allow for the global
helical nature of galactic magnetic fields that imprints on its structure and 
symmetries. The ambiguities, problems and limitations of the current approaches to 
GMF modelling have been discussed in \citet{Planck16} and by the IMAGINE consortium 
\citep{IMAGINE}.

A possible way to overcome heuristics in GMF modelling would be to extract the field 
structure from physical simulations of galaxy formation and evolution, which include all 
relevant processes of magnetic field generation and are specific to the MW. Such 
simulations of generic galaxies have led to important insights into various magnetic 
structures in spiral galaxies \citep[e.g.][]{Pakmor2017,2013MNRAS.432..176P}
but their resolution remains insufficient to capture galactic dynamo action, as it is
controlled by turbulent processes on scales less than 100\,pc. Moreover, the simulations 
would have to be constrained in a way flexible enough to reproduce an ever increasing set 
of observational data.

Here we propose a simple (that is, flexible, adjustable, analytic) and yet
realistic (i.e., based on relevant equations and observations) approach to model the
large-scale (mean) magnetic field in the disc and halo of the Milky Way and other
spiral galaxies. The model has been implemented as a publicly available software package
\galmag \citep{GALMAG_zenodo}, which 
can be used in the framework of Bayesian optimisers 
\citep[e.g.][]{Steininger2018, Steininger2018ASCL}

The text is organised as follows. In Section \ref{sec:BEq}, we lay out the modelling strategy, 
basic equations and fundamental assumptions. In Section \ref{sec:TDDS}, we describe the 
solutions for magnetic field in the disc and in Section \ref{sec:SDS}, for the halo.
In Section~\ref{sec:discuss} we discuss possible applications of this model to the interpretation 
of observations of synchrotron emission and Faraday rotation (Section~\ref{sec:radio}) as well 
some of the ways in which it could be used to model the evolution of galactic magnetic fields 
(Section~\ref{sec:Bevol}); possible extensions to this model are also discussed (Section~\ref{sec:EM}), 
whilst Section~\ref{galmag} introduces the publicly available \galmag software package that 
implements the model. Our results are summarised in Section~\ref{sec:final}.

\section{Basic equations}\label{sec:BEq}
A physically meaningful model of a galactic magnetic field has to rely on a clear
physical picture of its origin and maintenance, as well as on a specific galaxy model.
Mean-field dynamo action is the most plausible mechanism of generation and maintenance
of large-scale magnetic fields in spiral galaxies \citep{RSS88,Beck1996,BrSu05,Beck15}.
Therefore, we first explore the nature of magnetic structures compatible with dynamo 
action. However, magnetic fields observed in galaxies have many features that emerge
independently of the dynamo process. To allow for such features, we make our model rather
independent of the specific properties of dynamo-generated magnetic fields and use
solutions of the dynamo equations just as a convenient functional basis to parametrise
a wide class of magnetic structures. Furthermore, we discuss how an even more general
type of the governing equations can be used for these purposes.

The dynamo converts kinetic energy of random (turbulent) flows into magnetic 
energy. The mean helicity of the random flow, that emerges because of the overall rotation 
and stratification, leads to the generation of magnetic fields at scales much larger than 
the correlation scale of the random flow (this is described as the $\alpha$-effect). 
Differential rotation can accelerate the energy conversion by stretching the large-scale 
magnetic fields in the direction of the flow (the $\omega$-effect). 
The spatial scale and structure of the magnetic 
field are controlled by the mean-field transport coefficients, quantifying the averaged 
induction effects of the random flows, and the large-scale velocity shear rate.
The magnetic field structure also depends upon 
the geometric shape of the dynamo region (e.g. spherical, toroidal or flat).
When the magnetic field is weak, so that its effect on the velocity field is negligible, 
the dynamo leads to an exponential amplification of the large-scale magnetic field. 
As the Lorentz force becomes stronger, the field growth slows down and the system
gradually settles into a (statistically) steady state: the dynamo action saturates.

The mean-field dynamo equation has the form
\begin{equation}\label{MFD}
    \frac{\partial \vec{B}}{\partial t} =
    \nabla\times(\alpha\vec{B})
    +\nabla\times(\vec{V}\times\vec{B})
    + \etat\nabla^2\vec{B}\,,
\end{equation}
where $\vec{B}$ is the large-scale magnetic field, $\alpha$ and $\etat$
are the turbulent transport coefficients representing the mean induction
effects of the helical interstellar turbulence (the $\alpha$-effect)
and turbulent magnetic diffusion, respectively, and $\vec{V}$ is the large-scale
velocity field. The latter is dominated by differential rotation but can
also include galactic outflows (fountain or wind) and accretion.

Detailed reviews of the galactic dynamo and comprehensive references can be
found in \citet{RSS88, Beck1996, ShukurovMAND, ShukurovSubramanian2018}.
We present here a  very short overview of the theory with the number of
specific references reduced to a minimum.

The mean-field galactic dynamo equation has been solved under various 
approximations that cover a wide range of galactic environments. Our goal 
is to present a general class of physically-motivated magnetic field models that 
can be used to fit observations without the need to delve too deeply into the theory. 
Thus, we present a {parametrised} model with
the large-scale magnetic field in the form of an expansion over appropriate 
basis functions, specifically the modes of free decay which solve the diffusion equation in 
the disc and spherical geometries. The form of the magnetic structures obtained
is controlled by the choice of the expansion coefficients. The large-scale 
magnetic field in the model consists of a superposition of approximate solutions 
of the kinematic mean-field dynamo equations for a thin disc 
(Section~\ref{sec:TDDS}) and a spherical gaseous halo (Section~\ref{sec:SDS}).
Our use of the dynamo equations is less restrictive than it might seem since their
solutions can be used as a functional basis to represent a wide class of 
complex magnetic configurations, not necessarily produced by dynamo action, 
in terms of a small number of the expansion coefficients. Unlike representations
in terms of the Euler potentials \citep{Ferriere2014}, magnetic configurations
presented here can be helical. Another advantage of our approach is that all
variables and parameters of the model are either directly observable or related 
to observable quantities.

The approximate nature of the solutions that we use is due to the 
approximations adopted to solve the equations as well as the simple superposition 
of separate disc and halo solutions of the dynamo equation. Such a superposition 
is not quite consistent with the presumably non-linear nature of galactic
dynamos. However, the non-linear effects do not, plausibly,
affect the spatial distribution of the large-scale magnetic field too
strongly and the marginally stable dynamo eigenfunctions often provide a
satisfactory approximation for the non-linear solutions  \citep{CSSS14}.

The analytic solutions of the mean-field dynamo equations for the galactic discs
and halos presented here are obtained assuming that the disc is thin and the halo is
spherical. The thin-disc solutions are applicable at those distances to the galactic
centre $s$ where $h/s\lesssim0.1$, where $h$ is the scale height of the warm
ionised gas that is assumed to host the mean magnetic field.

\begin{table*}
\caption{Fiducial parameter values and input parameters of the \galmag code.
           \label{parameters}}
\centering
\begin{tabular}{llccr}
\hline\hline
Component  &Input parameter                                   &Equation          &Notation                       &Fiducial value\\
\hline
General    &Reference galactocentric radius                         &\eqref{eq:zs}     &$\rcyl_0$            &$8.5\kpc$ \\
\\
Disc       &Radius of the dynamo active disc                        &\eqref{eq:gamma'} &$\rcyl_{\rm d}$      &$17\kpc$ \\
           &Rotation curve                                          &                  &$V(\rcyl)$           &\citet{Clemens1985}\\
           &Dimensionless shear rate due to differential rotation   &\eqref{eq:RR}     &$\Rod$               &$-53$ \\
           &Dimensionless intensity of helical turbulence           &\eqref{eq:RR}     &$\Rad$               &$0.4$ \\
           &Disc shape                                              &\eqref{eq:scaleheight} &$h(\rcyl)$      &---\\
           &Disc scale height at $\rcyl=\rcyl_0$                    &\eqref{eq:zs}     &$h_0$                &$0.5\kpc$ \\
           &Azimuthal magnetic field strength at $\rcyl=\rcyl_0$    &	               &$B_{\rm d}$          &$-3\muG$ \\
           &Position of the first field reversal                    &                  &$\rcyl_{{\rm r}1}$   &$7\kpc$\\
           &Position of the second field reversal (Model B)         &                  &$\rcyl_{{\rm r} 2}$  &$12\kpc$\\
\\
Halo       &Radius of the dynamo active halo                        &\eqref{eq:rh}     &$r_{\rm h}$          &$15\,\kpc$ \\
           &Rotation curve                                          &\eqref{eq:Vhalo}, \eqref{eq:Vhalo2}  &$\bm{V}(\bm{r})$    & --- \\
           &Rotation curve turnover radius                          &\eqref{eq:Vhalo2} &$\rcyl_v$            &$3\,\kpc$ \\
           &Dimensionless shear rate due to differential rotation   &\eqref{eq:RRh}    &$\Roh$               &$-204$ \\
           &Dimensionless intensity of helical turbulence           &\eqref{eq:RRh}    &$\Rah$               &4.3/8.1 \\
           &Azimuthal magnetic field strength at $\rcyl=\rcyl_0$    &                  &$B_{\rm h}$          &$-0.5\muG$/$-0.01\,\muG$ \\
\hline
\end{tabular}
\end{table*}

\subsection{Symmetries of galactic magnetic fields}\label{sMFD}
It is convenient to introduce cylindrical polar coordinates $(s,\phi,z)$ with the
origin at the galactic centre and the $z$-axis parallel to the angular velocity 
$\vec{\Omega}$. Hence, $\vec{\Omega}=(0,0,\Omega)$ and $z=0$ at the galactic mid-plane. 
However, spherical coordinates $(r,\theta,\phi)$ are natural for the halo, with 
the polar axis $\theta=0$ aligned with the $z$-axis of the cylindrical frame and
the mid-plane (equator) at $\theta=\pi/2$.

Solutions of the dynamo equation are sensitive to the geometry of the dynamo region. 
In a thin disc, large-scale magnetic fields of even parity strongly dominate, that is,
$B_{s,\phi}(-z)=B_{s,\phi}(z)$ and $B_z(-z)=-B_z(z)$; this configuration 
corresponds to quadrupolar symmetry. Without dynamo action, quadrupolar magnetic fields 
in a thin disc decay slower than dipolar ones. As a result, for
realistic values of parameters, dipolar magnetic 
fields can be supported by the dynamo only in the central parts of the discs of 
spiral galaxies, $\lesssim1\kpc$. In a quasi-spherical halo, however, both odd 
(dipolar) and even (quadrupolar) magnetic fields can be maintained with almost equal 
ease, with $B_{r,\phi}(-z)=-B_{r,\phi}(z)$ and $B_\theta(-z)=B_\theta(z)$ for the 
dipolar symmetry and $B_{r,\phi}(-z)=B_{r,\phi}(z)$ and $B_\theta(-z)=-B_\theta(z)$ 
in the quadrupolar field (with $z=r\cos\theta$). Moreover, magnetic fields in the 
two halves of the halo, $z>0$ and $z<0$, can be disconnected by the disc, so that 
the symmetry of the magnetic field in the halo is only weakly constrained. The model 
proposed here provides freedom in selecting any symmetry of the magnetic field in the disc 
and the halo independently.

Despite deviations from axial symmetry, mainly associated with the spiral
pattern, galactic discs are sufficiently symmetric in azimuth that the 
axially-symmetric modes dominate the dynamo. Therefore,
deviations from axial symmetry in the large-scale magnetic field, however
strong they might be, can be included as distortions of a background
axially symmetric magnetic structure. In this paper, we mainly consider axially symmetric galaxies and axially symmetric magnetic fields and discuss extensions
to more general configurations in Section~\ref{sec:EM}.

\subsection{Boundary conditions}\label{BCs}
The simplest and best explored solutions of the mean-field dynamo equations
are obtained with the so-called vacuum boundary conditions, that is, under
the assumption that the electric current density outside the dynamo region
is negligible in comparison with any electric currents within it.
Equivalently, the magnetic diffusivity (inversely
proportional to electric conductivity) outside the dynamo region is
assumed to be much larger than within it. With vanishing electric current
density, $\nabla\times\vec{B}=\vec{0}$, the magnetic field is potential.

Neglecting external electric currents in the gaseous halo appears to be reasonable 
given the low density of intergalactic plasma.
Regarding galactic discs surrounded by a gaseous halo, it is important to note
that the magnetic diffusivity relevant to a large-scale magnetic field
is the \textit{turbulent} diffusivity. \citet{PSS93} argue that the turbulent
magnetic diffusivity in galactic haloes is about 50 times larger than in the
disc. This justifies the application of vacuum boundary conditions to the
\textit{large-scale} magnetic field at the disc surface as well. In other words,
we assume that the extension of the disc's magnetic field into the halo is a 
potential magnetic field that adds to the magnetic field
produced \textit{in situ} in the halo. 

In a thin disc, the vacuum boundary conditions have the form
$B_\phi(\pm h)=0$ and $B_s(\pm h)\approx0$, where $z=\pm h(\rcyl)$ is the disc
surface. The boundary condition for $B_\phi$ is exact whereas the accuracy of 
the boundary condition for $B_s$ is higher when the disc is thinner \citep{PSSS00,WSSS04}.
The potential magnetic field around the disc, that satisfies these boundary conditions,
is purely vertical. The vacuum boundary conditions for the halo are $B_\phi(r\h)=0$ and
$\nabla\times\vec{B}=\vec{0}$ at $r>r\h$, where $r\h$ is the halo radius.

\section{Magnetic field in the disc}\label{sec:TDDS}
\subsection{Rotation curve and disc thickness}
\begin{figure}
 \centering
 \includegraphics[width=0.95\columnwidth]{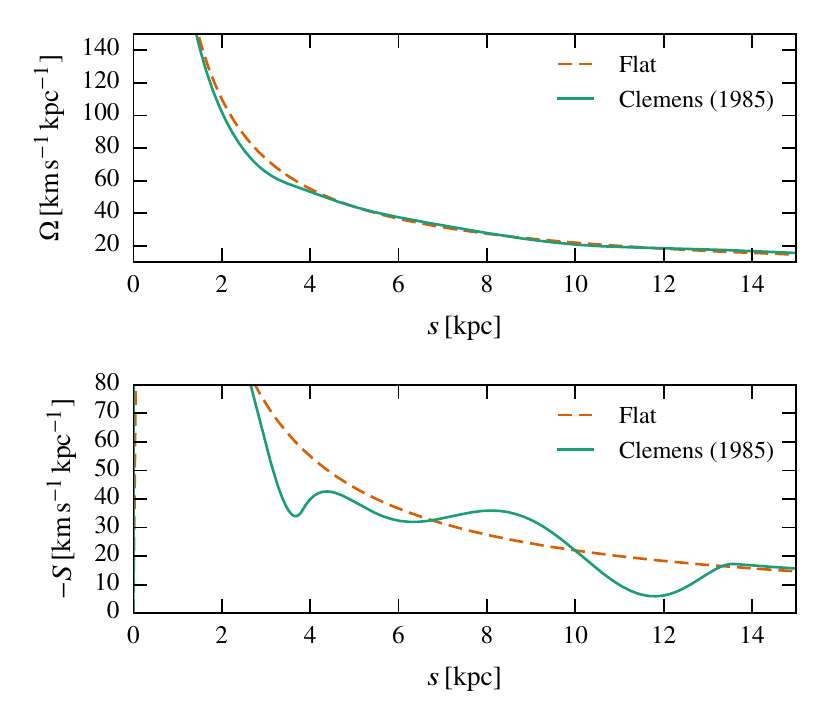}
 \caption{Two choices for the rotation curve discussed in the text (upper panel)
 and the corresponding velocity shear rate $S=\rcyl\dd\Omega/\dd\rcyl$ (lower panel): 
 the Milky Way rotation curve obtained from CO observations by \citet{Clemens1985}
 (solid) and a simpler rotation curve given by Eq.~\eqref{eq:simple_rotation_curve}
 (dashed).}
 \label{fig:rotation_curve}
\end{figure}

While the model can be applied to any galaxy, our choice 
of fiducial parameters is motivated by the Milky Way, with the disc rotation
curve of \citet{Clemens1985}. To illustrate the impact of the rotation curve on
the magnetic field, we also consider a flat rotation curve 
(i.e., the rotational speed is nearly independent of $\rcyl$ at large distances from 
the disc axis),
\begin{equation}\label{eq:simple_rotation_curve}
    V(\rcyl) = V_0\frac{1 -\exp(-\rcyl/\rcyl_*)}{1 -\exp(-\rcyl_0/\rcyl_*)}\,,
\end{equation}
where $\rcyl_0$ is a reference galactocentric distance defined below, $V_0$ is 
the rotation speed at $\rcyl=\rcyl_0$ and we take $\rcyl_*=250\pc$. 
The two rotation curves and the corresponding velocity shear rates, 
$S=\rcyl\dd \Omega / \dd \rcyl$, are shown in Fig.~\ref{fig:rotation_curve}.

The disc scale height is assumed to increase exponentially with the
cylindrical radius (a flared disc), 
\begin{equation}\label{eq:scaleheight}
  h(\rcyl)=h_0\exp\left(\frac{\rcyl-\rcyl_0}{\rcyl_\mathrm{h}}\right)\,,
\end{equation}
where we adopt a flaring length scale of $\rcyl_\mathrm{h}=5\kpc$, similar to that of 
the MW \HI disc, and $\rcyl_0$ is the Galactocentric distance of the Sun 
\citep{Kalberla2009}. In our fiducial model, we adopted the characteristic height
$h_0=0.5\kpc$, which is the scale height of the Lockman layer
\citep{Lockman1984,Dickey1990} near the Sun.

The radius of the dynamo-active part of the disc is chosen to be $\rcyl\di=17\kpc$, 
similar to the radius of the supernova distribution in the MW \citep{CaBh98}. 
The fiducial values of the parameters that appear in the model (which are also input
parameters for the \galmag code) are shown in Table~\ref{parameters}. Parameters of 
the halo are introduced in Section~\ref{PGH}.

\subsection{\label{ThDiDy}Thin-disc dynamos}
In terms of cylindrical coordinates $(\rcyl,\phi,z)$, the radial and azimuthal
components of the axisymmetric $\alpha^2\omega$-dynamo equation \eqref{MFD} can be
written as
\begin{align}
  \frac{\partial B_\rcyl}{\partial t} &=
    - \frac{\partial (\alpha B_\phi)}{\partial z}
    + \etat \frac{\partial^2 B_\rcyl}{\partial z^2}
    + \etat \frac{\partial}{\partial \rcyl}\left[
        \frac{1}{\rcyl}\frac{\partial (\rcyl B_\rcyl)}{\partial \rcyl}
        \right], \label{MFDc_radial}\\
  \frac{\partial B_\phi}{\partial t} &=
    S  B_\rcyl
    + \frac{\partial (\alpha B_\rcyl)}{\partial z}
    + \etat \frac{\partial^2 B_\phi}{\partial z^2}
    + \etat \frac{\partial}{\partial \rcyl}\left[
        \frac{1}{\rcyl}\frac{\partial (\rcyl B_\phi)}{\partial \rcyl}
        \right],\label{MFDc_azimuthal}
\end{align}
where $S=\rcyl\,\partial\Omega/\partial\rcyl$ is the velocity shear due to
differential rotation. We do not exhibit the equation for $B_z$ since, in a thin
disc, it decouples from the equations shown and can be solved
separately (equivalently, $B_z$ can be derived from $\nabla\cdot\vec{B}=0$). 

It is convenient to use dimensionless variables, denoted with tilde,
\begin{equation}\label{eq:zs}
\widetilde{s}=\rcyl/\rcyl_0 \qquad\text{and}\qquad
\widetilde{z}=z/h_0\,, \quad\text{with }\; h_0=h(\rcyl_0)\,,
\end{equation}
where $\rcyl_0$ is the reference cylindrical radius within the disc --
for example, $\rcyl_0=\rcyl_\odot\approx8.5\kpc$ is a convenient choice in the MW.
Using different length units across and along the disc allows us to make the disc
thinness explicit and quantified with the (small) aspect ratio
\begin{equation}\label{eps}
\epsilon=h_0/s_0\,.
\end{equation}
The large-scale velocity is measured in the units of a characteristic
rotational speed $V_0 =V(\rcyl_0)$,
\begin{equation}\label{Vom}
\widetilde{\vec{V}}=\frac{\vec{V}}{V_0}\,,
\quad
\widetilde{\Omega}=\Omega \frac{s_0}{V_0}\,.
\end{equation}
Velocity shear due to differential rotation is non-dimensionalised similarly,
\begin{equation}
\widetilde{S}=\frac{S}{S_0}\,,\text{ with }S_0=S(\rcyl_0)\,.
\end{equation}
The unit of time is the turbulent magnetic diffusion time across the disc.
With $\etat\di$ the turbulent magnetic diffusivity in the disc, the dimensionless 
time is
\begin{align}\label{dt}
\widetilde{t}&=t\etat\di/h_0^2\,.
\end{align}

The magnitude of the $\alpha$-effect can be estimated as
\begin{equation}\label{eq:an}
\alpha\simeq\min\left(l^2\Omega/h,v\right)\,,
\end{equation}
{where $l$ and $v$ are the turbulent scale and speed,}
and the corresponding fiducial value is used to non-dimensionalise $\alpha$:
\begin{equation}\label{ad}
  \widetilde{\alpha}=\frac{\alpha}{\alpha_0}\,,
\quad
  \alpha_0=\frac{l^2 V_0}{h_0 s_0}\,.
\end{equation}
The magnitude of $\alpha$ cannot exceed $\alpha=v$ because it is a measure of the
helical part of the turbulent flow speed, hence $\alpha/v$ cannot exceed unity.
This limit is usually important only in the central parts of galaxies
(where the thin-disc approximation does not apply anyway). Equation~\eqref{eq:an} 
gives the magnitude of $\alpha$ and its dependence on $\rcyl$ through the 
variations of $\Omega$ and $h$ with $\rcyl$. It is expected that $\alpha$ is 
an odd function of $z$. \citet{GZER08a,GEZR08b} and \citet{BGE15} confirm this 
and discuss the dependence of $\alpha$ on $z$ in numerical simulations of the 
supernova-driven interstellar medium. We adopt a factorised form for $\alpha(\rcyl,z)$,
\begin{equation}\label{asz}
\widetilde{\alpha}(s,z)=\frac{\widetilde{\Omega}(s)}{\widetilde{h}(s)}\az(z)\,, 
\qquad\text{where}\qquad a(-z)=-a(z)\,,
\end{equation}
where we assume that $l$,
in Eqs~\eqref{eq:an} and \eqref{ad}, is independent of $\rcyl$. The model 
can be generalised  straightforwardly to more general forms of $\alpha(s,z)$.

When galactic outflow is neglected, the dynamo is fully characterised by two 
dimensionless
control parameters that quantify the intensity of the mean magnetic induction
due to helical turbulence and differential rotation, respectively:
\begin{equation}\label{eq:RR}
\Rad=\frac{h_0\alpha_0}{\etat\di}=\frac{l^2 V_0}{\rcyl_0 \etat\di}\,,
\quad
\Rod=\frac{h_0^2 S_0}{\etat\di}\,,
\end{equation}
where the subscript `d' refers to the disc (similar parameters are defined
slightly differently in the halo -- see Section~\ref{sec:SDS}).

In terms of dimensionless variables, Eqs.~\eqref{MFDc_radial} and
\eqref{MFDc_azimuthal} reduce to
\begin{align}
  \deriv{B_\rcyl}{\widetilde{t}} &=
    -\Rad \deriv{(\widetilde{\alpha}B_\phi)}{\widetilde{z}}
    + \deriv{^2 B_\rcyl}{\widetilde{z}^2}
    + \epsilon^2 \deriv{}{\widetilde{\rcyl}}
    \left[
  \frac{1}{\widetilde{\rcyl}}\deriv{(\widetilde\rcyl B_\rcyl)}{\widetilde{\rcyl}}
        \right], \label{eq:MFDc_radial}\\
  \deriv{B_\phi}{\widetilde{t}} &=
    \Rod \widetilde{S} B_\rcyl
    + \Rad \deriv{(\widetilde{\alpha} B_\rcyl)}{\widetilde{z}}
    + \deriv{^2 B_\phi}{\widetilde{z}^2}
     + \epsilon^2 \deriv{}{\widetilde{\rcyl}}
\left[\frac{1}{\widetilde{\rcyl}}\deriv{(\widetilde{\rcyl}B_\phi)}{\widetilde{\rcyl}}
        \right].\label{eq:MFDc_azimuthal}
\end{align}
It is now clear that the solutions are fully determined by $\Rad$, $\Rod$, 
$\widetilde{S}(\rcyl)$, $\widetilde{\alpha}(\rcyl)$ and $\epsilon$.

In a thin disc, the magnetic field distribution along $z$ is established over a time 
scale $h^2/\etat\di\simeq5\times10^8\yr$ which is $\epsilon^{-2}=(\rcyl_0/h_0)^2$ 
times shorter than the time scale at which the radial distribution evolves, 
$\rcyl_0^2/\etat\di$. Because of this difference, the radial derivatives in 
Eqs.~\eqref{eq:MFDc_radial} and \eqref{eq:MFDc_azimuthal} have $\epsilon^2$ 
as a factor. Therefore, the magnetic field distribution in a thin disc can be
represented as a local solution (at a given galactocentric distance $s$),
$\vec{b}(z;s)$, modulated by an envelope $Q(s)$. The local solution
also depends on $s$ since $\alpha$, $S$ and $h$ vary with $s$, but this
variation is parametric. It is convenient to normalise the local solution
to unit surface magnetic energy density, $\int_{-h}^h |\vec{b}|^2\,\od z=1$,
at all values of $s$: then $Q(s)$ represents magnetic field strength
at the galactocentric radius $s$ (an \textit{envelope} of the local solutions).
Thus, asymptotic solutions of Eqs.~\eqref{eq:MFDc_radial} and 
\eqref{eq:MFDc_azimuthal} for $\epsilon\ll1$ have the form
\begin{equation}\label{eq:aproxDisc}
 \vec{B}(s,z,t) = \exp(\Gamma t)\, Q(\rcyl) \, \vec{b}(z;\rcyl)\,.
\end{equation}
The magnetic field varies exponentially with time at a rate $\Gamma$ in the
kinematic dynamo stage. In a saturated thin-disc dynamo, the solution has
a similar form but with $\Gamma=0$ \citep{PSS93}. The local solution is 
discussed in Section \ref{sec:disk_local}, whereas Section \ref{sec:disk_radial} 
presents the radial solution $Q(\rcyl)$.

To simplify the notation, we use exclusively the dimensionless variables
with the tilde suppressed in the remaining part of this section unless
otherwise stated.

\subsection{Local solutions}\label{sec:disk_local}

Governing equations for the local solution,
$\vec{b}=\exp{(\gamma t)}\,(b_\rcyl,b_z)$, follow from 
Eqs.~\eqref{eq:MFDc_radial} and \eqref{eq:MFDc_azimuthal} when we put
$\epsilon=0$:
\begin{align}
\gamma(\rcyl) b_\rcyl&
=-\Rad\deriv{}{z}\left[\alpha(\rcyl,z) b_\phi\right]+\deriv{^2 b_\rcyl}{z^2}\,,
        \label{eq:locals}\\
\gamma(\rcyl) b_\phi &= \Rod S(\rcyl) b_\rcyl
+\deriv{^2 b_\phi}{z^2}+ \Rad \deriv{}{z}\left[\alpha(\rcyl,z) b_\rcyl\right]\,.
     \label{eq:MFDc_azimuthal_alt}
\end{align}
To allow for the disc flaring, we introduce the following new variables:
\begin{equation}
  \hz = \frac{z}{h(\rcyl)}\,,\quad
  \widehat{b}_\rcyl=\frac{b_\rcyl}{\Rad\Omega(\rcyl)} \quad
  \text{and}\quad
  \widehat{\gamma} = \gamma(\rcyl)\,h^2(\rcyl) \,.
\end{equation}
Since $\Rad \ll \Rod$ at $\rcyl\gtrsim1\kpc$ in most spiral galaxies, we can omit the term proportional to 
$\Rad$ in Eq.~\eqref{eq:MFDc_azimuthal_alt}, thus obtaining the $\alpha\omega$-dynamo 
approximation:
\begin{align}
    \widehat{\gamma}(\rcyl)\widehat{b}_\rcyl&
            =-\frac{\partial}{\partial\hz}\left[ a(z) b_\phi\right]
             +\frac{\partial^2 \widehat{b}_\rcyl}{\partial\hz^2}\,,
             \label{eq:local}\\
    \widehat{\gamma}(\rcyl) b_\phi &= \Dlocal(\rcyl)
    \widehat{b}_\rcyl+ \frac{\partial^2 b_\phi}{\partial\hz^2}\,,\label{local}
\end{align}
where  $\az(z)$, the $z$-dependent part of $\alpha(\rcyl,z)$, is defined in Eq~\eqref{asz}
and $\Dlocal(\rcyl)$, the
\textit{local dynamo number}, includes the radial variation of the dynamo
parameters:
\begin{equation}\label{eq:Dlocal}
  \Dlocal(\rcyl) = \Rad \Rod\, \Omega(\rcyl) S(\rcyl) h^2(\rcyl)\,.
\end{equation}
It is also convenient to introduce the local (in galactocentric radius) values 
of $\Rad$ and $\Rod$,
\begin{equation}\label{Raloc}
\mathcal{R}_\alpha(\rcyl)=\Rad\Omega(\rcyl)\,,
\quad
\mathcal{R}_\omega(\rcyl)=\Rod S(\rcyl)h^2(\rcyl)\,,
\end{equation}
so that $\Dlocal=\mathcal{R}_\alpha\mathcal{R}_\omega$.
The normalisation condition for the local solution reduces to
\begin{equation}\label{intb}
\int_{-1}^1 \left(|\widehat{b}_s|^2+|{b}_\phi|^2\right)\,\od\hz=1\,,
\end{equation}
where we have neglected $\widehat{b}_z$ because this component of magnetic 
field is, on average, weaker than the other two.
To this order in $\epsilon$, $b_z$ does not enter the equations for $b_\rcyl$ 
and $b_\phi$ and can be solved for separately. The result can be shown to be 
identical to that obtained from $\nabla\cdot\vec{B}=0$. 

The vacuum boundary conditions are
\begin{equation}\label{bcd}
\widehat{b}_\rcyl=b_\phi=0 \quad \text{at}\quad \hz=\pm1\,. 
\end{equation}
Since $\az(z)$ is an odd function of $z$, solutions of Eqs.~\eqref{eq:local}
and \eqref{local} split into two independent classes, even and odd in $z$
(or quadrupolar and dipolar, respectively). These can be distinguished using
the symmetry conditions at the galactic mid-plane,
\begin{align}
\frac{\partial \widehat{b}_\rcyl}{\partial\hz} &=
\frac{\partial b_\phi}{\partial\hz}= b_z= 0\quad\text{at\ } z=0
\quad\text{(even\ parity)},
\label{eq:24}
\\
\widehat{b}_s&=b_\phi= \deriv{b_z}{\hz}=0 \quad\text{at\ } z= 0
\quad\text{(odd\ parity)}.
\label{eq:quad_sym}
\end{align}
Approximate solutions for both parity families can be obtained in the form of an
expansion over the free-decay modes obtained as solutions of
Eqs.~\eqref{eq:local} and \eqref{local} for $a(z)=\Dlocal(\rcyl)=0$.
The procedure is discussed in detail by \citet{ShSo08} and \citet{CSSS14}, and
here we only provide the results.

\begin{figure}
 \centering
 \includegraphics[width=0.8\columnwidth]{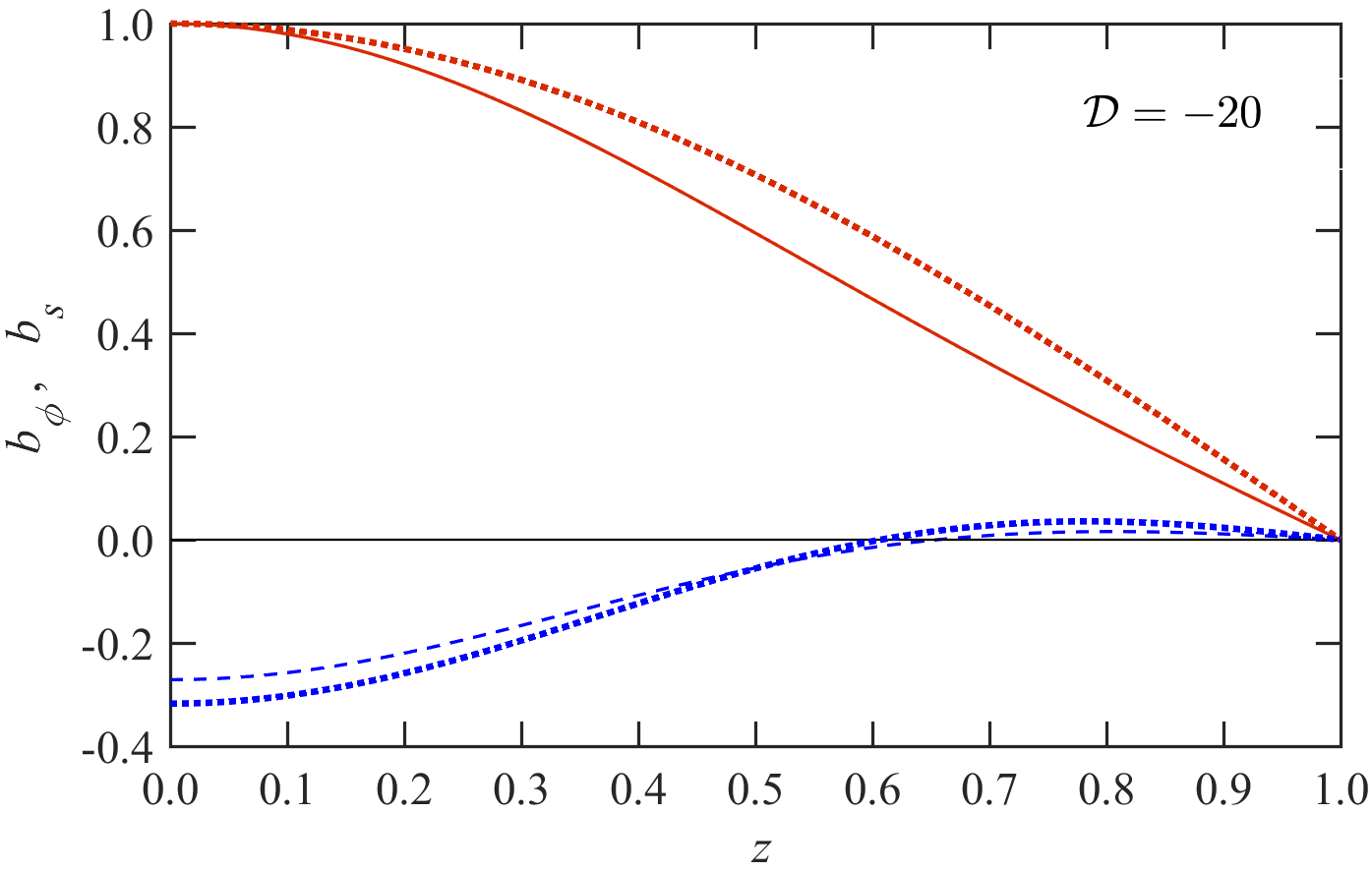}\\
 \includegraphics[width=0.8\columnwidth]{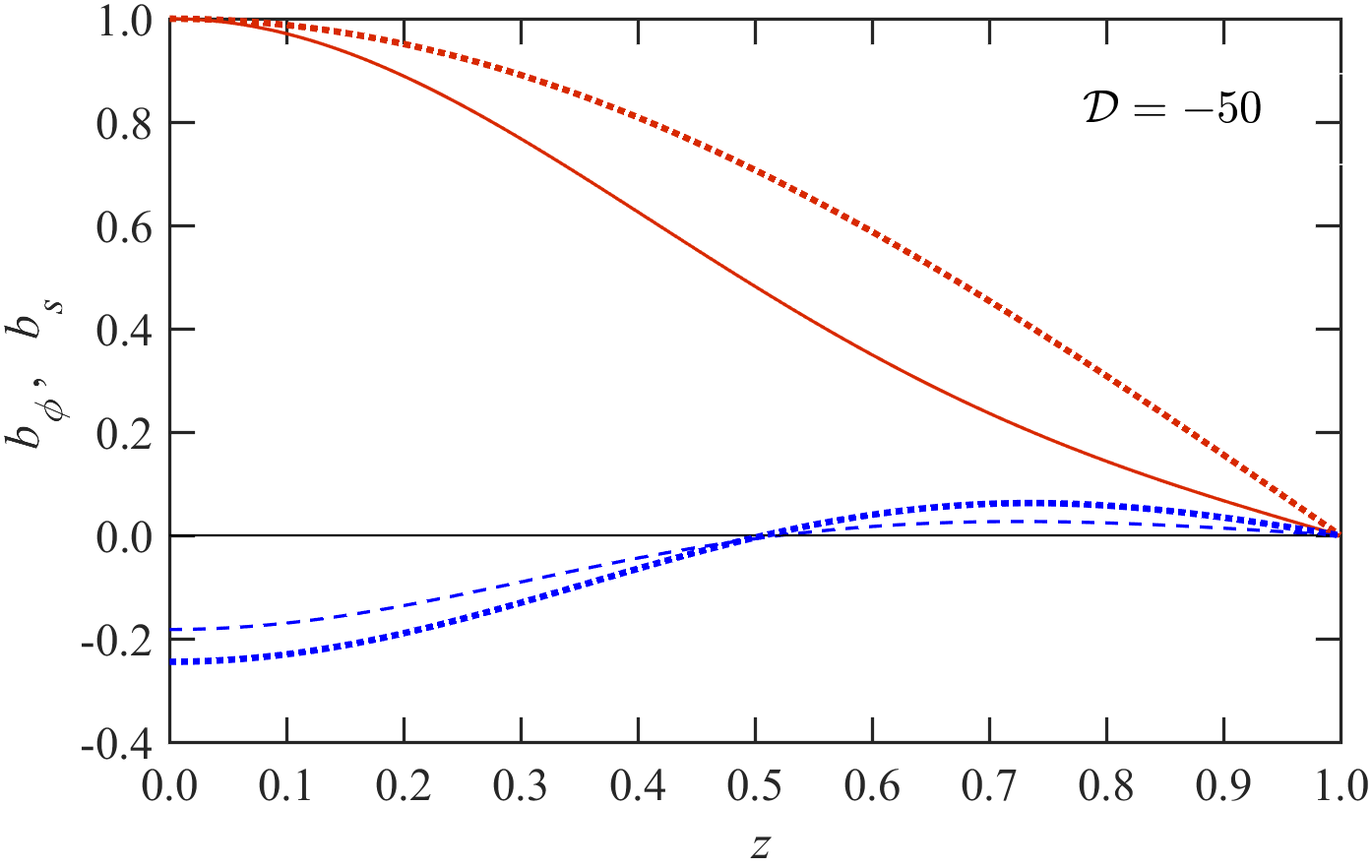}\\
 \includegraphics[width=0.75\columnwidth]{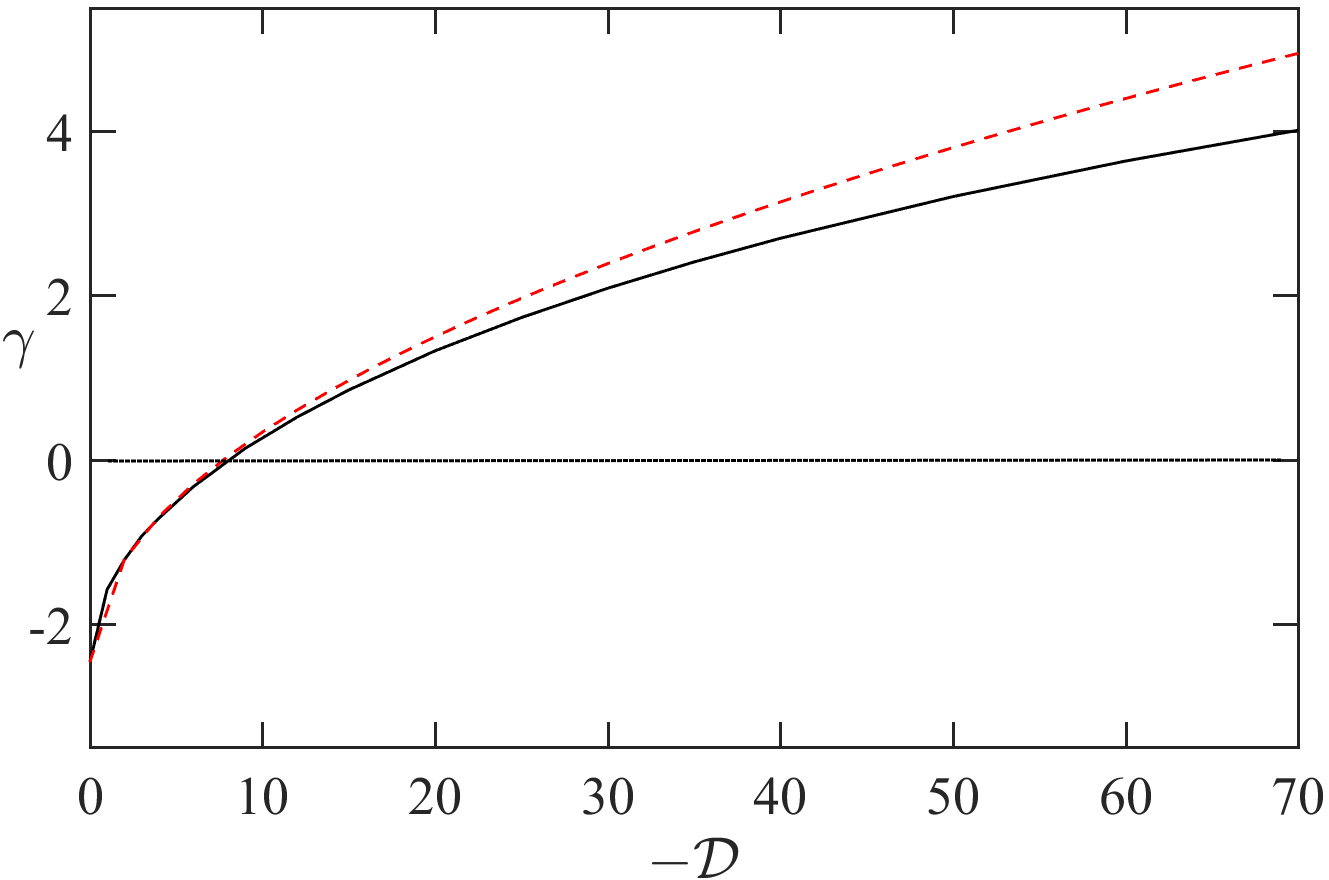}\\
 \caption{Comparison of the local quadrupolar eigenfunctions for $\Dlocal=-20$ 
 (upper panel) and $-50$ (middle panel) obtained from numerical solution of the 
 local equations \eqref{eq:local} and \eqref{local} with the approximate 
 eigenfunctions 
 \eqref{eq:local_bs_qua} and \eqref{eq:local_bphi_qua}: 
 $b_\phi$ from the numerical solution are shown solid (red) and 
 $b_s$, dashed (blue); their approximate counterparts are shown with dotted 
 curves of the matching colour. The eigenfunctions are normalised to 
 $b_\phi(0)=1$. The bottom panel shows the numerical (black, solid) and 
 approximate, Eq.~\eqref{eq:local_gamma_qua} (red, dashed), solutions for the 
 local growth rate as a function of the local dynamo number.
 \label{local_s}
 }

\end{figure}

From Eqs.~\eqref{eq:local}--\eqref{eq:24}, we obtain the following
approximate solutions of quadrupolar symmetry, for 
${a(\hz)=\sin(\pi\hz/2)}$:
\begin{align}
b_\rcyl(z;\rcyl) &\approx \mathcal{R}_\alpha{(\rcyl)}
K_0(\rcyl)
  \left[\cos\frac{\pi z}{2h(\rcyl)} 
  +\frac{3\sqrt{-\Dlocal(\rcyl)}}{4\pi^{3/2}}
                        \cos\frac{3\pi z}{2h(\rcyl)}\right]
                        \,,\label{eq:local_bs_qua}\\
b_\phi(z;\rcyl) &\approx -2K_0{(\rcyl)} \sqrt{-\frac{\Dlocal(\rcyl)}{\pi}}
                    \cos\frac{\pi z}{2h(\rcyl)}
                        \,,\label{eq:local_bphi_qua}\\
 \gamma(\rcyl) &\approx \frac{1}{h^2(\rcyl)}\left[
    -\frac{\pi^2}{4}+\frac{1}{2}\sqrt{-\pi \Dlocal(\rcyl)}
    \right]
                    \,,
 \label{eq:local_gamma_qua}
\end{align}
where
\begin{equation}
 K_0{(\rcyl)} = \left[1 - \frac{4 \Dlocal(s)}{\pi} - \frac{9 \Dlocal(s)}{16 \pi^{3}}\right]^{-1/2}
\end{equation}
is the normalization factor obtained using Eq.~\eqref{intb}.
These solutions are shown in Fig.~\ref{local_s}. 
Equation~\eqref{eq:local_gamma_qua} provides the critical value of the 
local dynamo number required for the local amplification and maintenance 
of the magnetic field: $\gamma\geq0$ for $\Dlocal\leq\Dlocal\crit\approx-\pi^3/4\approx-8$.
Other choices of the functional form of $a(\hz)$ lead to
slightly different values of the critical dynamo number (e.g. $-11$ for
$a=z$), but the difference hardly has any practical consequences.

In the odd-parity solutions, the term proportional to $(-\Dlocal)^{1/2}$ in
$b_\rcyl$ vanishes for $\az(\hz)=\sin(\pi\hz/2)$. Therefore, it is more convenient 
to use a similar solution with $\az(z)=\hz$ that satisfies the symmetry condition 
\eqref{eq:quad_sym}, here written to the lowest order in $(-\Dlocal)^{1/2}$:
\begin{align}
b_\rcyl(z;\rcyl) &\approx K_1{(\rcyl)}
	\mathcal{R}_\alpha{(\rcyl)} \sqrt2\sin\frac{\pi z}{h(\rcyl)}\,,	\label{eq:local_bs_dip}\\
b_\phi(z;\rcyl) &\approx -2 K_1{(\rcyl)}
		\sqrt{-\Dlocal(\rcyl)} \sin\frac{\pi z}{h(\rcyl)}\,,		\label{eq:local_bphi_dip}\\
\gamma(\rcyl)   &\approx \frac{1}{h^2(\rcyl)}\,,					\label{eq:local_gamma_dip}
\end{align}
with 
\begin{equation}
 K_1{(\rcyl)} = \left[1 - 4 \Dlocal(s)\right]^{-1/2}\,.
\end{equation}
The dipolar
modes can be excited, $\gamma\geq0$, for $\Dlocal\leq-2\pi^4\approx-195$, a 
threshold much higher than for the quadrupolar modes. This is true
for any plausible form of $a(\hz)$ and explains the 
predominance of quadrupolar magnetic fields in thin discs. 

The local solutions are derived, formally, for $|\Dlocal|\ll1$ but they remain 
reasonably accurate for $|\Dlocal|$ as large as about 50 or more \citep{2JCBS14}. 
We compare in Fig.~\ref{local_s} the local quadrupolar eigenfunctions obtained 
from numerical solution of Eqs.~\eqref{eq:local} and \eqref{local} with the 
approximate solutions \eqref{eq:local_bs_qua}--\eqref{eq:local_gamma_qua} for 
$\Dlocal=-20$ and $-50$, values typical of the main parts of spiral 
galaxies (see Fig.~\ref{fig:local_gamma}).
 
The local eigenfunctions presented above are approximate solutions of the 
mean-field dynamo equations. However, the functional basis of these solutions, the 
free-decay modes, can be used as a complete functional basis to represent any 
magnetic field configuration.  
The free-decay modes are solutions of Eqs.~\eqref{eq:local} and 
\eqref{local} with $a(z)=0$ and $\Dlocal(\rcyl)=0$, so that the equations decouple 
and can easily be solved.
Normalised as in Eq.~\eqref{intb}, the dipolar and quadrupolar free-decay eigenfunctions 
and eigenvalues (identified with superscripts $d$ and $q$, respectively) 
have the following respective forms with $n=1,2,3,\ldots$:
\begin{equation}\label{fdd}
\vec{b}_n^{(\text{d})}=\begin{pmatrix}\sin(\pi n\hz)\\0 \end{pmatrix}\,,
\quad
\vec{b}_n^{(\text{d})\prime}=\begin{pmatrix}0\\\sin(\pi n\hz)\end{pmatrix}\,,
\quad
\widehat{\gamma}_n^{(\text{d})}=-\pi^2 n^2\,,
\end{equation}
\begin{align}\label{fdq}
\vec{b}_n^{(\text{q})}
&=\begin{pmatrix}\cos\left[\pi\left(n-\tfrac12\right)\hz\right]\\0 \end{pmatrix}\,,
\  
\vec{b}_n^{(\text{q})\prime}
=\begin{pmatrix}0\\\cos\left[\pi\left(n-\tfrac12\right)\hz\right] \end{pmatrix}\,,\\
\widehat{\gamma}_n^{(\text{q})}&=-\pi^2 \left(n-\tfrac12\right)^2\,.\label{fdqg}
\end{align}
The free-decay disc modes are double degenerate as two orthogonal modes
of the same parity (distinguished by prime) correspond to each eigenvalue. 

\subsection{Radial solution}\label{sec:disk_radial}
When Eq.~\eqref{eq:aproxDisc} is substituted into Eqs.~\eqref{eq:MFDc_radial} and
\eqref{eq:MFDc_azimuthal}, and Eqs.~\eqref{eq:locals} and \eqref{eq:MFDc_azimuthal_alt}
are allowed for, equations for both $B_\rcyl$ and $B_\phi$ reduce to the same equation
for $Q(s)$, that is,
\begin{equation}\label{eq:Qs}
\epsilon^2\frac{\partial}{\partial \rcyl}\left[
\frac{1}{\rcyl}\frac{\partial}{\partial \rcyl}(\rcyl Q)\right]
+[\gamma(s)-\Gamma]Q = 0\,,
\end{equation}
where $\gamma(s)$ is the local growth rate obtained as a part of the local 
solution, \eqref{eq:local_gamma_qua} or \eqref{eq:local_gamma_dip}.
This equation applies to both even and odd local solutions, and to both
$\alpha^2\omega$ and $\alpha\omega$-dynamos,
that is, with and without the term proportional to $\Rad$ in Eq.~\eqref{eq:MFDc_azimuthal_alt}.
The boundary conditions adopted are
\begin{equation}\label{bcQ}
Q(0)=Q(\rcyl\di)=0\,. 
\end{equation}
The condition at $\rcyl=0$ follows from axial symmetry, whereas that at the outer 
boundary of the dynamo-active region $\rcyl=\rcyl\di$ is adopted for the sake of 
simplicity. 

\begin{figure}
 \centering
 \includegraphics[width=0.95\columnwidth]{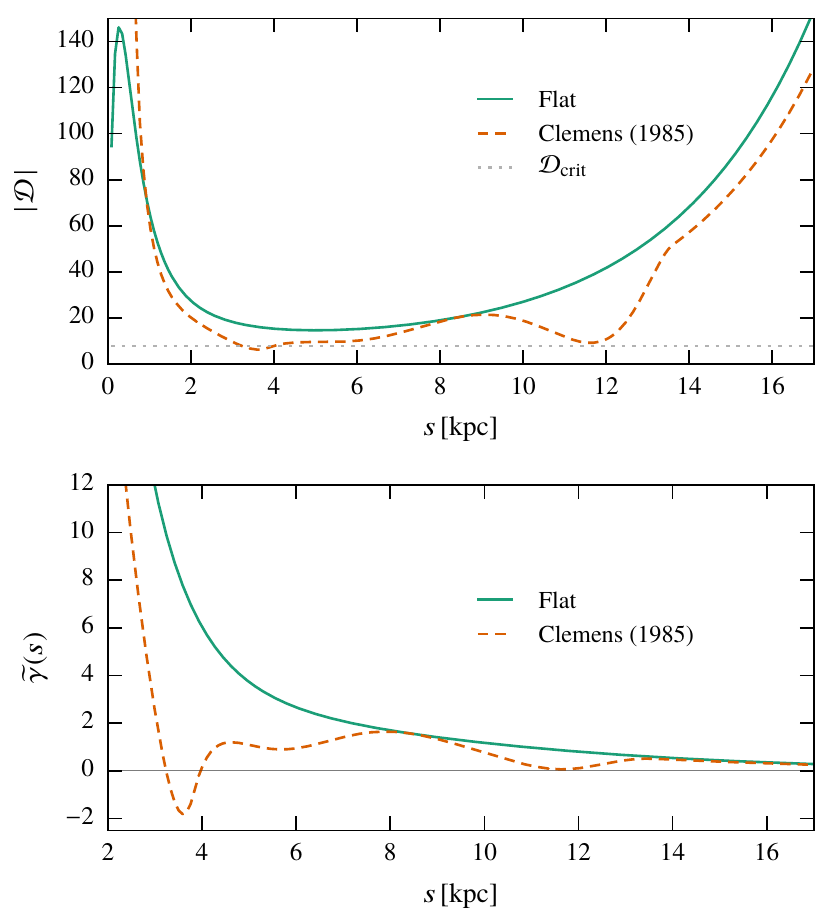}
 \caption{
 Radial profiles of the local dynamo number (upper panel) and the local
 growth rate of the quadrupolar solutions
 (lower panel) for the flat rotation curve \eqref{eq:simple_rotation_curve} (solid) and for 
 the MW rotation curve of \citet{Clemens1985} (dashed). In the upper panel, the critical dynamo 
 number $\Dlocal\crit\approx-\pi^3/4$ is shown dotted for reference.
}
 \label{fig:local_gamma}
\end{figure}

\begin{figure}
    \centering
\includegraphics[width=0.97\columnwidth]{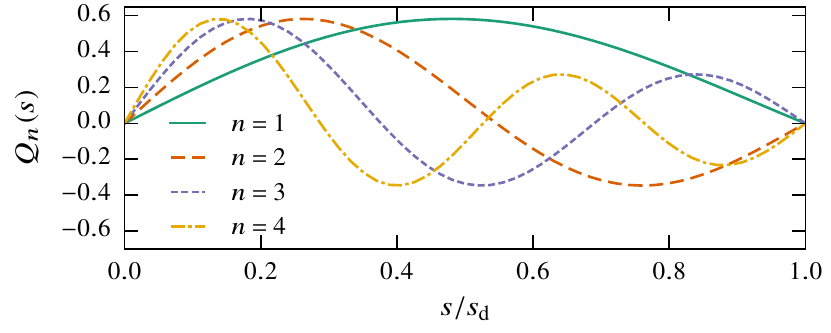}
    \caption{
Four leading eigenmodes $Q_n(\rcyl)$ of the radial equation \eqref{eq:Qs} given by 
Eq.~\eqref{eq:Qn_sol}.
}
    \label{fig:disc_details}
\end{figure}

When $\Omega(s)$ and $h(s)$ are known, the global growth rate $\Gamma$ and
the radial distribution of the magnetic field strength $Q(\rcyl)$
can be obtained by solving Eq.~\eqref{eq:Qs} numerically.
However,  for our present purposes it is more useful to obtain an approximate analytical 
solution for $Q(s)$. This allows faster computations at the cost of
an additional approximation. When the magnetic field model is used in Bayesian
analyses, the computation speed is of primary importance.

To obtain a simple analytical solution of Eq.~\eqref{eq:Qs}, consider $\gamma$ to be a 
constant, $\gamma=\gamma_0$, within the disc radius, $\rcyl<\rcyl\di$, and zero outside,
\begin{equation}
 \gamma(\rcyl)= \begin{cases}
    \gamma_0\, ,& 0\leq \rcyl<\rcyl\di \,, \\
    0       \,,& \rcyl\geq \rcyl\di \,.
    \end{cases}\label{eq:gamma'}
\end{equation}
\noindent A suitably averaged value of $\gamma(\rcyl)$ within the disc can be adopted for 
$\gamma_0$. The 
relevance of this approximation depends on the specific case, particularly the rotation 
curve and the rate of the disc flaring, as shown in Fig.~\ref{fig:local_gamma}. The inner 
parts of the disc, $\rcyl\lesssim1\kpc$, should be disregarded since the thin-disc
approximation is not applicable there. At $\rcyl\gtrsim1\kpc$, the approximation
$\gamma(\rcyl)=\const$ does not appear unreasonable, especially for the
Milky Way rotation curve, as shown in the lower panel
of Fig.~\ref{fig:local_gamma}.

For $\gamma(\rcyl)$ of Eq.~\eqref{eq:gamma'}, equation \eqref{eq:Qs} with boundary 
conditions \eqref{bcQ} can be solved to yield the eigensolutions
\begin{align}
 Q_n(\rcyl)&= J_1\left(k_n \rcyl/\rcyl\di\right)\,,
 \qquad n=1,2,3,\ldots\,,\label{eq:Qn_sol}\\
\label{Gamma_n}
\Gamma_n &= \gamma_0 -\epsilon^2 k_n^2 \,,
\end{align}
where $J_1(x)$ is the standard Bessel function (of order one) and 
$k_n\approx~3.83, 7.02, 10.17,\ldots$ are its zeros,
thus, $J_1(k_n)=0$.
The first four modes of $Q_n(\rcyl)$ are shown in Fig.~\ref{fig:disc_details}.
Independently of the form of $\gamma(\rcyl)$, the lowest radial mode, $Q_1(s)$, is 
sign-constant but $Q_n(\rcyl)$ has $n-1$ zeros, and thus the scale of variation of 
$Q_n(s)$ decreases with $n$. This feature of the solution is responsible 
for the reversals of the large-scale magnetic field discussed in Section~\ref{FRAGR}.

The evolving radial distribution of magnetic field is obtained as a superposition
of the eigensolutions of Eq.~\eqref{eq:Qs}:
\begin{equation}\label{QCnQn}
Q(\rcyl,t)=\sum_{n=1}^\infty C_n\e^{\Gamma_n t} Q_n(\rcyl)\,,
\end{equation}
where, in the context of dynamo models, the coefficients $C_n$ are determined by the 
initial conditions. Alternatively, the eigenfunctions $Q_n(\rcyl)\vec{b}(z;\rcyl)$ can be used as 
the basis functions to represent a given magnetic field with $\exp(\Gamma_n t)$ absorbed
into $C_n$. The set of radial eigenfunctions $Q_n$ is 
complete, so that any radial distribution of magnetic field can be represented in this 
form. The only constraint on the form of the solution is due to the fact that the set of
the local solutions $\vec{b}$ is incomplete. Nevertheless, a wide class of magnetic
field distributions along $z$ can be represented as a superposition of various quadrupolar 
and dipolar local modes, so that the lack of the functional completeness of the local 
solutions is not likely to be restrictive in practice. Otherwise, the complete set of
local free-decay modes $\vec{b}_m^{\text{(q,d})}(z;\rcyl)$ of Eqs.~\eqref{fdd}--\eqref{fdqg} 
can be used instead of the local dynamo solutions
to construct a more general expansion over a 
complete set of basis functions of the form
\[
\vec{B}(\rcyl,z)=\sum_{m=1}^\infty \sum_{n=1}^\infty C_{mn}Q_n(\rcyl)\vec{b}_m^{\text{(q,d})}(z;\rcyl)\,,
\]
which can be used to parametrise (in terms of the coefficients $C_{mn}$) an arbitrary
magnetic field that does not need to be a solution of the dynamo equations.

\subsection{Vertical magnetic field}
\label{sec:disk_vertical}

The axially symmetric vertical component of the magnetic field, $B_z$, can be obtained from
\begin{equation}\label{divB}
 \nabla\cdot\vec{B}=\frac{1}{\rcyl}\deriv{}{\rcyl}( \rcyl B_\rcyl) + \deriv{B_z}{z} =0\,,
\end{equation}
using
$ 
 B_\rcyl(\rcyl, z) = b_\rcyl(z; \rcyl) Q(\rcyl)
$ 
from Eq.~\eqref{eq:local_bs_qua} or~\eqref{eq:local_bs_dip},
and Eq.~\eqref{eq:Qn_sol}.
We have assumed that $\gamma_0=\const$ to derive the simple form  Eq.~\eqref{eq:Qn_sol},
implying $\Dlocal=\const$. It is therefore justifiable to neglect
the dependence of $\Dlocal$, $\Omega$ and $h$ on $\rcyl$ when differentiating
$B_\rcyl$ in \eqref{divB} and only retain the dependence of $Q$ on $\rcyl$  (this 
simplification can easily be relaxed if required). By virtue of linearity,
Eq.~\eqref{divB} can be solved for each radial mode $n$ separately: 
\begin{align}
-\deriv{B_z^{(n)}}{z}&=\frac{1}{\rcyl}\deriv{}{\rcyl}(\rcyl B_\rcyl^{(n)})=
  \mathcal{R}_\alpha K_0 C_n\frac{1}{\rcyl} \oderiv{}{\rcyl}\left[\rcyl J_1(k_n\rcyl/\rcyl_\text{d})\right]\nonumber\\
  &\times \left(\cos\frac{\pi z}{2h(\rcyl)}+
  \frac{3}{4\pi^{3/2}}\sqrt{-\Dlocal(\rcyl)}
                            \cos\frac{3\pi z}{2h(\rcyl)}\right)\,.
\end{align}
Then, for the quadrupolar parity,
\begin{align}\label{Bzint}
  B_z^{(n)}&  = -{\mathcal{R}_\alpha K_0 C_n}
   \frac{k_n}{\rcyl_\text{d}} 
   J_0\left(\frac{k_n \rcyl}{\rcyl_\text{d}}\right)\nonumber\\
   &\mbox{}\quad\times\int_0^z\!\left(\cos\frac{\pi z'}{2h}
       +\frac{3\sqrt{-\Dlocal}}{4\pi^{3/2}}\cos\frac{3\pi z'}{2h}\right) \dd z'
\end{align}
where the dependence of $\mathcal{R}_\alpha$ $\Dlocal$, $h$ and $K_0$ on $\rcyl$ should
be allowed for.

\subsection{Magnetic field in the disc}

We can now collect the solutions from Sections~\ref{sec:disk_local}, \ref{sec:disk_radial}
and \ref{sec:disk_vertical} to write the approximate solution of the dynamo equation
(evolving or at a fixed time) for a thin disc as a sum of $N_r$ radial eigenmodes,
\begin{equation}\label{Bsumn}
 \begin{pmatrix} B_\rcyl \\ B_\phi \\ B_z \end{pmatrix} 
 = \sum_{n=1}^{N_r} C_n
 \begin{pmatrix}   B_\rcyl^{(n)} \\ B_\phi^{(n)} \\ B_z^{(n)} \end{pmatrix}\,,
\end{equation}
where, for the quadrupolar symmetry,  
\begin{align}\label{eq:Bdisc_R}
B_\rcyl^{(n)} &= K_0(\rcyl) \mathcal{R}_\alpha(\rcyl) 
J_1(k_n \rcyl/\rcyl_\text{d}) \nonumber\\
           &\quad\times\left[ \cos\frac{\pi z}{2 h(\rcyl)}
                       +\frac{3\sqrt{-\Dlocal(\rcyl)}}{4\pi^{3/2}}
                               \cos\frac{3\pi z}{2h(\rcyl)}\right],\\
\label{eq:Bdisc_phi}
B_\phi^{(n)} &= -2\sqrt{-\frac{\Dlocal(\rcyl)}{\pi}} K_0(\rcyl)
	J_1(k_n\rcyl/\rcyl_\text{d})\cos\frac{\pi z}{2h(\rcyl)}\,,\\
\label{eq:Bdisc_z}
B_z^{(n)} &= -\frac{2k_n h(\rcyl)}{\pi\rcyl_\text{d}} K_0(\rcyl) \mathcal{R}_\alpha(\rcyl)
  J_0(k_n\rcyl/\rcyl_\text{d}) \nonumber\\
  &\quad \times \left[\sin\frac{\pi z}{2h(\rcyl)}
    +\frac{\sqrt{-\Dlocal(\rcyl)}}{4\pi^{3/2}}
    \sin\frac{3\pi z}{2h(\rcyl)}\right],
\end{align}
and similarly for the dipolar symmetry. The local dynamo
parameters $\Dlocal(\rcyl)$ and $\mathcal{R}_\alpha(\rcyl)$ are defined in 
Eqs.~\eqref{eq:Dlocal} and \eqref{Raloc}, whereas an example of the form of 
$h(\rcyl)$ is given by Eq.~\eqref{eq:scaleheight}.
In what follows (and in the \galmag code), we normalise each eigenmode so that each 
expansion coefficient represents the strength of the corresponding part of the
magnetic field at the reference radius $\rcyl_0$, that is, $|C_n| = |B^{(n)}(\rcyl_0,0)|$.

The coefficients $C_n$ can be chosen to fix the strength and to reproduce any radial distribution of the
magnetic field. For instance, when used to approximate the magnetic field observed in the 
disc, $C_n$ are used to fit the observations to a desired accuracy. When used to simulate a
growing magnetic field, its evolution is introduced through
$C_n = C_n^{(0)}\mathrm{e}^{\Gamma_n t}$, where the initial values $C_n^{(0)}$ 
at $t=0$ are obtained from the similar expansion for the seed magnetic field. Galactic 
evolution can be included by an appropriate time variation of $\Dlocal(\rcyl)$, 
$\Omega(\rcyl)$, $h(\rcyl)$ and $\rcyl\di$.

\begin{figure*}
 \centering
\includegraphics[width=\textwidth]{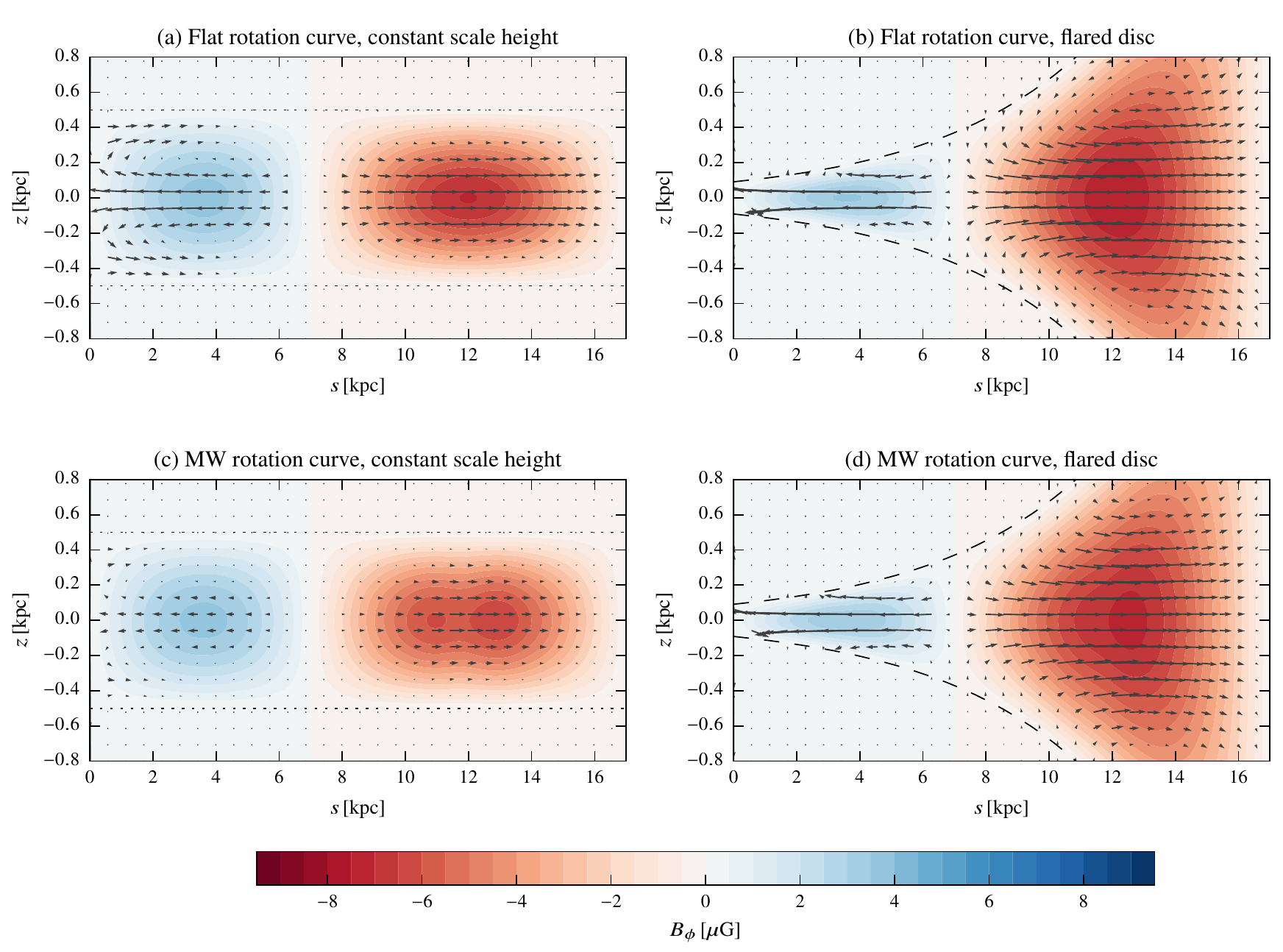}
 \caption{Sensitivity of the magnetic field structure to the rotation curve
 and disc flaring, illustrated with magnetic fields constructed using fiducial parameters
 and $(C_1,C_2) = (4.6\muG, -1.6\muG)$. The strength of the azimuthal component of the
 magnetic field is colour coded, while arrows indicate the direction and  strength of the 
 poloidal magnetic field. Details of the disc models are indicated above each  frame and the 
 disc scale height is indicated with a dotted line on the left and a dashed line on the right. 
 Strong differential rotation near the galactic centre leads to a strong magnetic field, 
 especially in a flat disc. In a flared disc, which is thinner at small $\rcyl$,  the field 
 strength near the galactic centre is reduced but the outer region has a stronger field.}
 \label{fig:scaleheight}
\end{figure*}

\subsection{Dynamo parameters}
In order to use the solution presented above, the dimensionless dynamo control 
parameters $\Rad$ and $\Rod$ have to be specified. We adopt the reference radius
$\rcyl_0=\rcyl_\odot\approx 8.5\kpc$ and, correspondingly, $V_0=220\kms$ and 
$S_0=-35\kms\kpc^{-1}$ as obtained from the rotation curve. 
An estimate of the turbulent magnetic diffusivity widely used in
various applications derives from the mixing length theory,
\begin{equation}\label{eq:beta}
\etat=\tfrac13 lv\,,
\end{equation}
where $l$ and $v$ are the turbulent scale and speed, respectively. With 
$l\simeq50\pc$ \citep[see][and references therein]{HSSFG17} and 
$v\simeq10\kms$ \citep[e.g.][]{MLK04}, we have
$\etat\simeq5\times10^{25}\cm^2\s^{-1}$. Equations~\eqref{eq:beta} and 
\eqref{eq:RR} then yield
\begin{equation}
\Rad = 0.39\,,\qquad \Rod = -53\,.
\end{equation}
A summary of parameters used by \galmag and their fiducial values can be
found in Table~\ref{parameters}.

\begin{figure*}
 \centering
 \includegraphics[width=\textwidth]{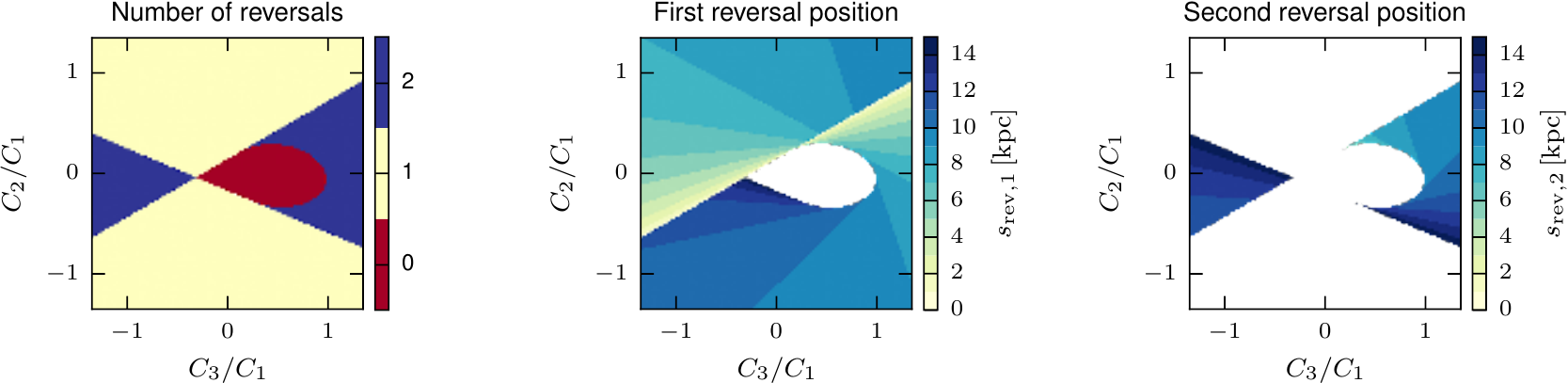}
 \caption{Number and positions of magnetic field reversals along the disc radius, for a magnetic field constructed by a superposition of the first three
fastest growing eigenfunctions, $Q_n(\rcyl)$ with $n=1,2,3$, for the fiducial
 choice of parameters shown in Table~\ref{parameters}. The left-hand panel shows 
 the number of the field reversals (colour coded as indicated with the colour bar on the
 right of the panel) for various ratios of the expansion coefficients  $C_n/C_1$. The 
 middle and right-hand panels show the galactocentric distances of the inner and outer
 reversal, $s_\text{rev,1}$ and $s_\text{rev,2}$ respectively, colour coded with the 
 colour bar to the right of each panel.  White colour indicates the absence of the 
 corresponding reversal.
          }
 \label{fig:reversals}
\end{figure*}
\begin{figure*}
 \centering
 \includegraphics{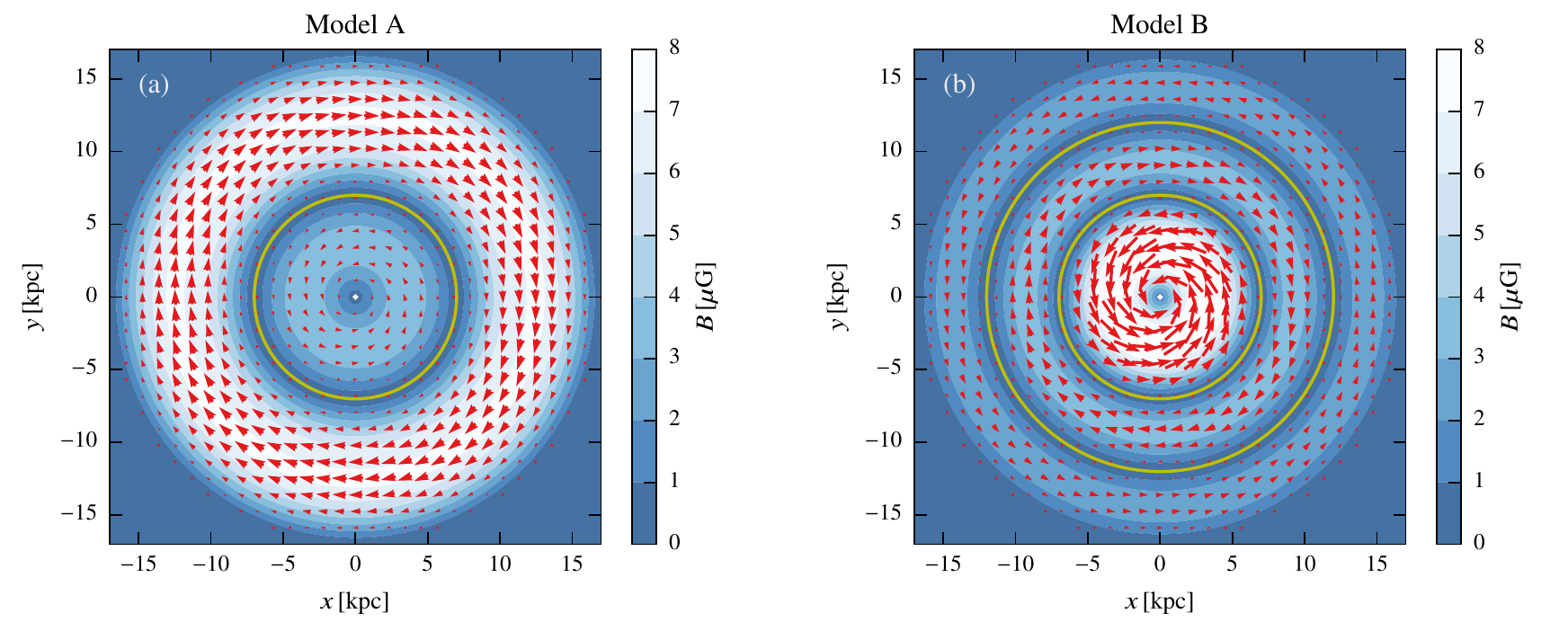}
 \medskip
 \includegraphics{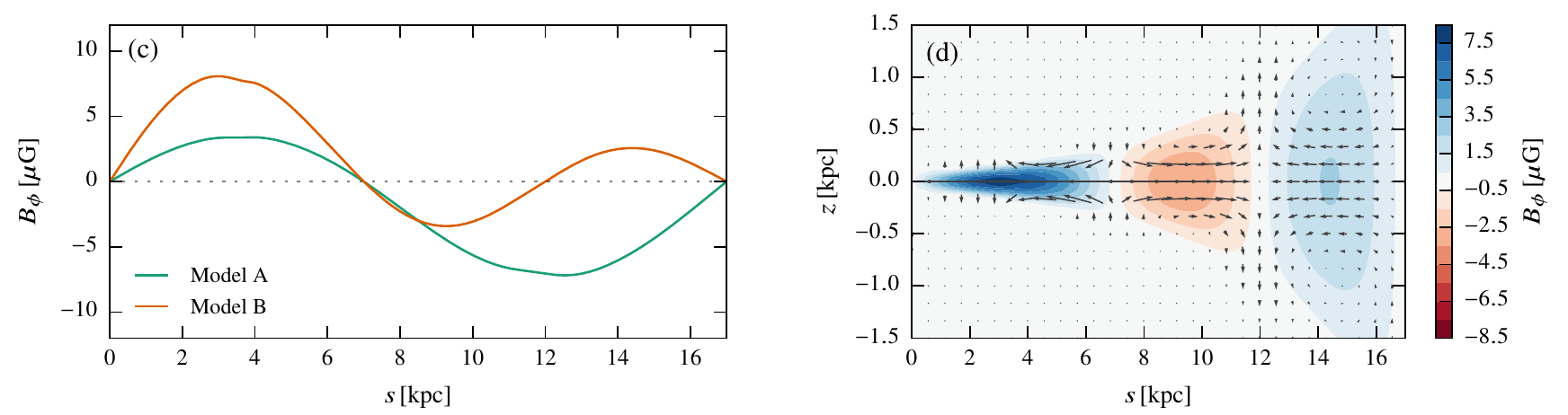}
 \caption{Two examples of the disc magnetic field with radial reversals. In the top 
 panels, the field strength in the galactic mid-plane is shown colour coded with
 arrows showing the direction and strength of the magnetic field projected onto the 
 mid-plane. Model A, shown in panel (a), has a reversal at $\rcyl=7\kpc$ and
 corresponds to the same field as in the bottom right panel of Fig.~\ref{fig:scaleheight}.
 Model B, shown in panel (b), has the same parameters except for having two reversals
 at $\rcyl=7$ and $12\kpc$. Panel (c) shows the azimuthal magnetic field in 
 the  two models. In panel (d), the vertical cross-section of Model B is shown, with the
 magnitude of azimuthal magnetic field indicated with colour and arrows showing the projection
 of the magnetic field onto the $(xz)$-plane.
           \label{fig:disc_field}}
\end{figure*}

\subsection{The role of the disc flaring and details of the rotation curve}

Figure~\ref{fig:scaleheight} shows the vertical cross section of magnetic field in the 
disc, constructed using the first two radial eigenmodes with 
$(C_1,C_2) = (4.6\muG,-1.5\muG)$.
These distributions have an absolute maximum of $|\vec{B}|$ at large~$\rcyl$. 
This happens because the local dynamo number of Eq.~\eqref{eq:Dlocal} 
increases with galactocentric distance in an exponentially flared disc,
$\Dlocal\propto \Omega S h^2\propto s^{-2}\exp(2\rcyl/\rcyl_\text{h})$ 
for a flat rotation curve, $\Omega\propto S\propto s^{-1}$. If $h$ indeed increases
with $s$ faster than $s^2$, the outer parts of galactic discs can have relatively strong 
magnetic fields at early stages of magnetic field growth when these kinematic solutions 
apply. Such distributions may occur in young and evolving galaxies. Nonlinear dynamo 
effects eventually limit the local magnetic field strength to a value related to 
equipartition between magnetic and turbulent kinetic energies,
$B^2\approx 4\pi\rho v^2(\Dlocal/\Dlocal_\text{cr}-1)
     \propto \rcyl^{-2} \exp(2\rcyl/\rcyl_\text{h} -\rcyl/\rcyl_\rho)$
assuming that $\rho\propto \exp(-\rcyl/\rcyl_\rho)$. The radial profile of magnetic 
field strength then depends on the relation between the radial length scales of the gas
density and disc thickness. The number density of \ion{H}{i} in the MW has
$\rcyl_\rho\simeq3\kpc$ \citep{Kalberla2009} which is close to 
$\rcyl_\text{h}/2\simeq2.5\kpc$. Therefore, we cannot exclude the possibility that the magnetic field
remains strong in the outer MW. The effective boundary of the dynamo active region is 
then determined by the rapid increase of the local dynamo time scale 
$\gamma^{-1}(\rcyl)\simeq h^2(\rcyl)/\etat\di$ with $\rcyl$ in a flared disc:
this time scale exceeds $10^{10}\yr$ where $h\gtrsim1\kpc$ for 
$\etat\di=5\times10^{25}\cm^2\s^{-1}$, and the growth of magnetic field becomes
practically negligible. For $h(\rcyl)$ given by Eq.~\eqref{eq:scaleheight}, this
happens at $\rcyl\gtrsim11\kpc$ provided $\etat\di$ is independent of $\rcyl$. When 
the magnetic field is stronger in the outer parts of the 
disc, the boundary condition $Q(\rcyl\di)=0$ may be too restrictive. We have 
considered solutions with $\partial(\rcyl Q)/\partial\rcyl|_{\rcyl=\rcyl\di}=0$
to confirm that the magnetic field distribution in the main part of the disc is not 
significantly affected but the outer field maximum becomes more pronounced.

The increase of the local dynamo number with distance from the galactic 
centre enhances the large-scale magnetic field in the outer parts of a galactic
disc in either the kinematic or saturated dynamo. As a result, the magnetic 
field energy density may decrease with radius slower than other energy 
densities in the interstellar medium, as suggested by \citet{Beck07}.

\subsection{\label{FRAGR}Field reversals along the galactocentric radius}
Magnetic field reversals can be reproduced naturally in the model because the radial 
eigenfunction $Q_n(\rcyl)$ has $n-1$ zeros. With an appropriate selection of the expansion
coefficients $C_n$, any desired number of reversals located at any prescribed positions
can be produced. While the strength of the magnetic field is controlled by the 
magnitudes of the expansion coefficients $C_n$, the number of reversals and their 
positions along the radius are controlled by the ratios of the coefficients,
for instance, $C_n/C_1$. Since $Q_2(\rcyl)$ has one zero while $Q_3(\rcyl)$ has two zeros,
retaining only the two leading terms in the expansion \eqref{Bsumn} allows us to obtain 
a magnetic field with one radial reversal, whereas in order to have two reversals,
$Q_3(\rcyl)$ needs to be included. To help selecting the coefficients as required to
obtain a magnetic field that has a given number of reversals at desired positions, we 
present Fig.~\ref{fig:reversals} where the ranges of $C_2/C_1$ and $C_3/C_1$  that produce
one or two reversals can be read off the left-hand panel. The desired positions of the 
reversals can be converted into the coefficient ratios using the middle and right-hand 
panels. Similar diagrams can be constructed for any number of reversals if required.

Figure~\ref{fig:disc_field} illustrates the structure of axisymmetric magnetic fields 
that have radial reversals. Model~A, with a reversal at $\rcyl=7\kpc$, represents
the same field as in the bottom right panel of Fig.~\ref{fig:scaleheight}.
Model~B, constructed using the three leading eigenfunctions, has reversals at 
$\rcyl = 7$ and $12\kpc$. In both models, the magnetic field is normalised so that
the azimuthal magnetic field, shown in Fig.~\ref{fig:disc_field}c, has the strength 
$B_{\phi}^{(\mathrm{d})}|_{\rcyl=\rcyl_0,\, z=0}=-3\muG$ in the mid-plane $z=0$ 
at the reference radius $\rcyl=\rcyl_0=8.5\kpc$. 

Figure~\ref{fig:disc_field}d shows the vertical cross-section (a meridional
plane) of the magnetic structure in Model~B to demonstrate that, in a thin disc, the
horizontal magnetic field components dominate over the vertical field, 
$|B_\phi|>|B_\rcyl|\gg |B_z|$, only on average. Locally, and especially near the 
reversals and the disc axis, the vertical magnetic field dominates. This is a direct 
consequence of the solenoidality of magnetic field: if $B_\phi$ and $B_\rcyl$ are weak, 
$B_z$ needs to be stronger to ensure that $\nabla\cdot\vect{B}=0$. The dominance of
$B_\phi$ over $B_\rcyl$ is less general in origin: this is a consequence of the 
stretching of the radial magnetic field by differential rotation, and the larger is the 
velocity shear the larger is the ratio $|B_\phi/B_\rcyl|$ and the smaller is the 
magnitude of the magnetic pitch angle $p=\arctan(B_\rcyl/B_\phi)$.

\section{Magnetic field in the halo}\label{sec:SDS}
\begin{figure*}
    \includegraphics[width=\textwidth]{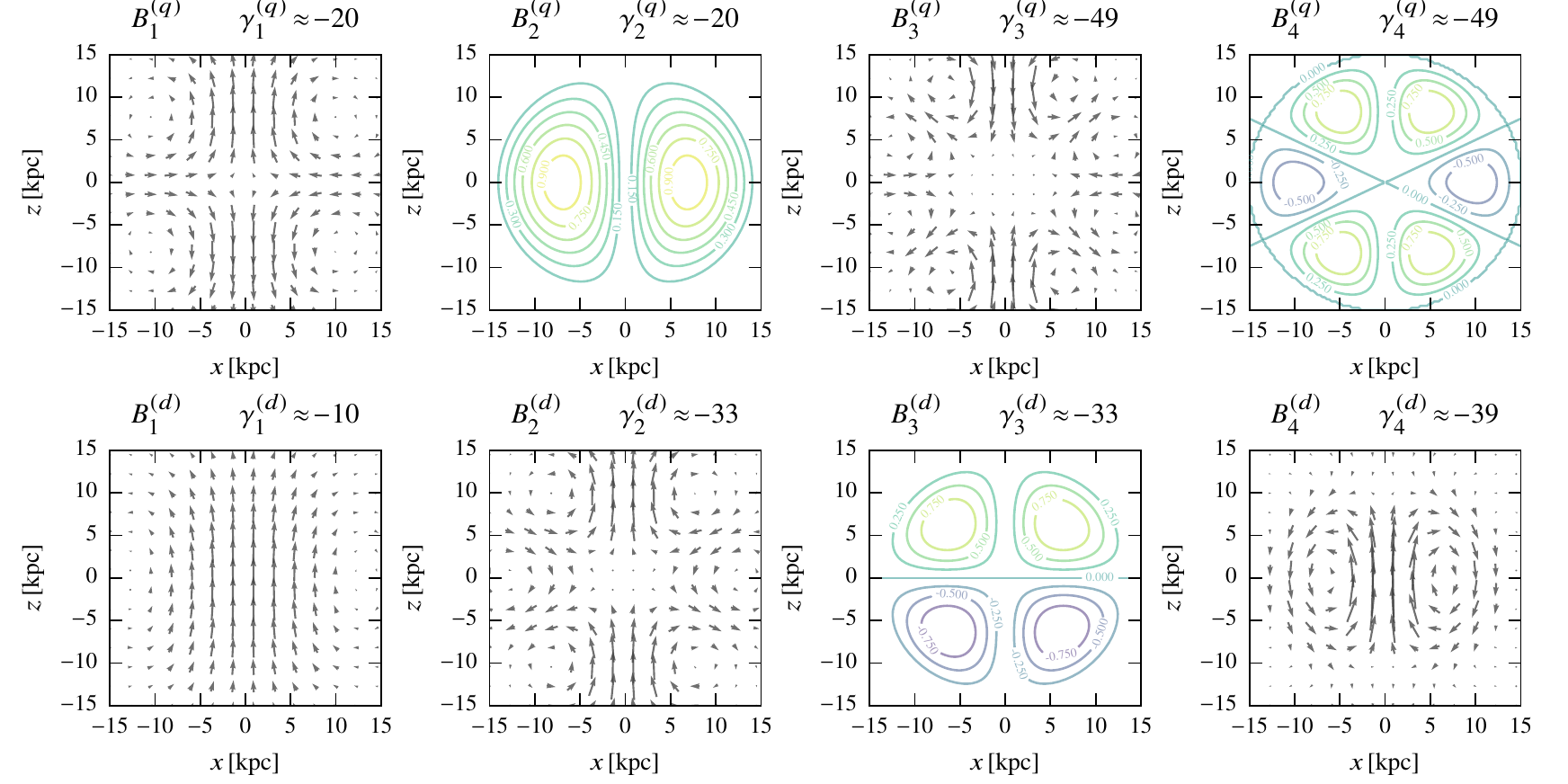}
\caption{Eight spherical free-decay eigenfunctions $\vec{B}_i$ of the smallest decay
rates. Each mode is either purely toroidal or purely poloidal.
The top row shows the modes symmetric with respect to the mid-plane $z=0$ (quadrupolar
modes), while the modes in the bottom row are anti-symmetric (dipolar). For the poloidal
modes, arrows represent the projection of the magnetic field on the $(xy)$-plane. For the
toroidal modes, contours show the strength of the azimuthal component of the magnetic field 
with the normalisation \eqref{hnorm}. The decay rate of each mode $\gamma$ is shown at the top of 
each panel. \label{fig:free_decay} }
\end{figure*}

Magnetic field in the spherical halo is obtained as the perturbation solution of the
mean-field dynamo equation with free-decay eigenfunctions as the unperturbed solutions. 
This approach is similar to that employed to obtain the local disc solution in 
Section~\ref{sec:disk_local}.

In the spherical halo, it is convenient to use spherical coordinates 
$(r,\theta, \phi)$. As for the disc, we define convenient dimensionless variables 
distinguished by the tilde: spherical radius and time are measured in the units of 
the halo radius $r\h$ and the corresponding magnetic diffusion time, respectively,
\begin{equation}\label{eq:rh}
\widetilde{r}=r/r\h \qquad \text{and}\qquad\widetilde{t}=t\etat\h/r\h^2\,,
\end{equation}
with $\etat\h$ the turbulent magnetic diffusivity in the halo. The velocity field and 
the $\alpha$-coefficient are normalised as 
\begin{equation}\label{alVh}
\widetilde{\alpha}=\alpha/\alpha\h\,,\qquad\widetilde{V}=V/V\h\,,
\end{equation}
where $\alpha\h$ is the $\alpha$-coefficient at the north pole, $(r,\theta)=(r\h,0)$,
and $V\h$ is the equatorial rotation velocity at the boundary $(r,\theta)=(r\h,\pi/2)$.

In terms of the dimensionless variables, the mean-field dynamo Equation \eqref{MFD}
reduces to
\begin{equation}
 \frac{\partial \vec{B}}{\partial \widetilde{t}} = \Rah \widetilde{\nabla}\times
 (\widetilde{\alpha}\vec{B})
 + \Roh \widetilde{\nabla}\times(\widetilde{\vec{V}}\times\vec{B})+\widetilde{\nabla}^2\vec{B}\,,
 \label{eq:MFDhalo}
\end{equation}
where we defined, analogously to Eq.~\eqref{eq:RR}, the dynamo parameters
\begin{equation}\label{eq:RRh}
\Rah=r\h\alpha\h/\etat\h\,,
\qquad
\Roh=-r\h V\h/\etat\h\,.
\end{equation}
To avoid excessively heavy notation, we suppress the tilde on the dimensionless variables
and work exclusively with dimensionless variables unless otherwise stated.

Solutions of Eq.~\eqref{eq:MFDhalo}, growing or decaying at a rate $\Gamma$, are 
sought in the form of an expansion
\begin{equation}\label{expan}
 \vec{B} = \exp(\Gamma t)\sum^N_{i=1}a_i
 \vec{B}_i(\vec{r})
\end{equation}
in the free-decay modes $\vec{B}_i$ which are obtained as solutions of 
Eq.~\eqref{eq:MFDhalo} with $\Rah=\Roh=0$,
\begin{equation}
 \nabla^2\vec{B}_i=\gamma_i \vec{B}_i\,.
 \label{eq:free_decay}
\end{equation}
where $\gamma_{i} < 0$ is the rate of exponential decay of the mode $\vec{B}_i$.
Outside the halo, an electromagnetic vacuum is assumed, implying a potential magnetic 
field, $\nabla\times\vec{B}_i=\vec{0}$. The boundary conditions that ensure a
continuous matching, at the halo boundary $r=1$, of the interior magnetic field to a
potential exterior magnetic field that decays at infinity as the point dipole 
(the lowest magnetic multipole) are given by \citep{M78}
\begin{equation}
\left[\vec{B}_i\right]=0 \ \text{at} \ r=1\,, 
\qquad \vec{B}_i=\mathcal{O}\left(r^{-3}\right)
\ \text{for} \ r\rightarrow\infty,
\label{bcs}
\end{equation}
where the square brackets denote the jump of the corresponding quantity. 

The spatial form and decay rates of the 
spherical modes of free decay are derived in Appendix~\ref{ap:free_decay}; here 
we briefly discuss their properties. The free decay modes form a complete, orthonormal 
set of basis functions (related to spherical harmonics), each either purely poloidal 
(comprising the field components $B_r$ and $B_\theta$) or purely toroidal (consisting of 
$B_\phi$ alone). They can be divided into two classes based on their symmetry about the 
equator $\theta=\pi/2$:  the symmetric modes are quadrupolar (indicated with 
superscript `q') whereas the anti-symmetric modes have a dipolar symmetry (superscript
`d'). Their analytic forms can be found in Appendices~\ref{sec:free_symm} and 
\ref{sec:free_anti}, respectively. Figure~\ref{fig:free_decay} shows the structure of 
the four free-decay modes of each symmetry that have the largest $\gamma_i$.

\begin{figure*}
 \centering
\includegraphics[width=0.9\columnwidth]{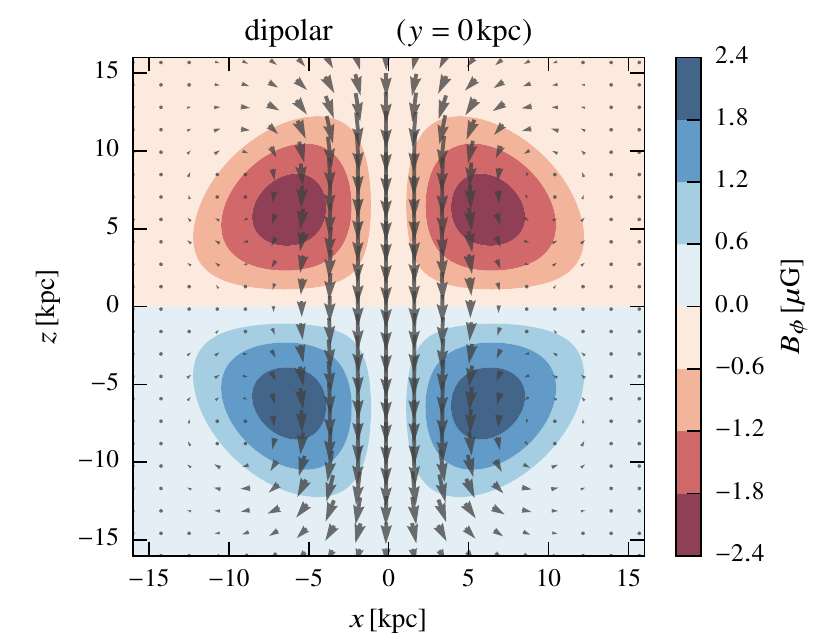}\qquad
\includegraphics[width=0.9\columnwidth]{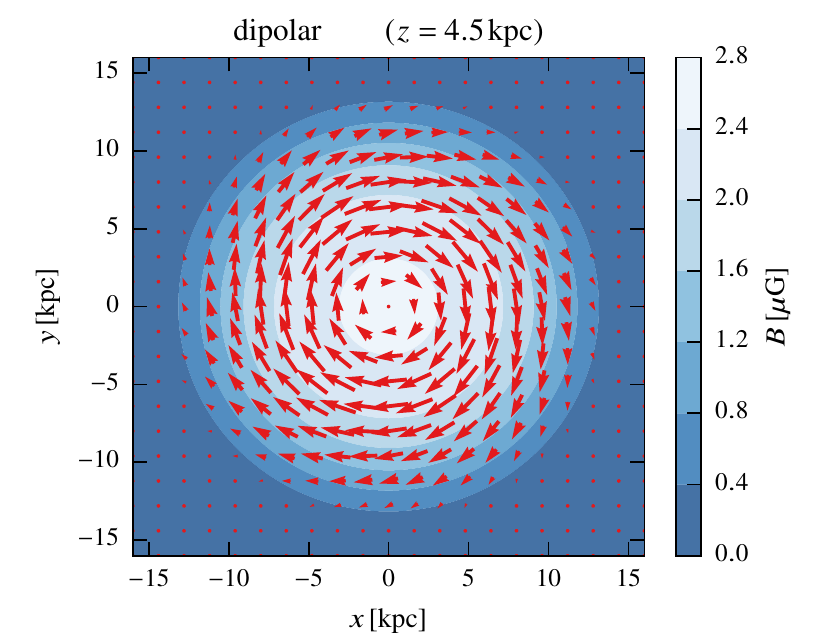}\\
\includegraphics[width=0.9\columnwidth]{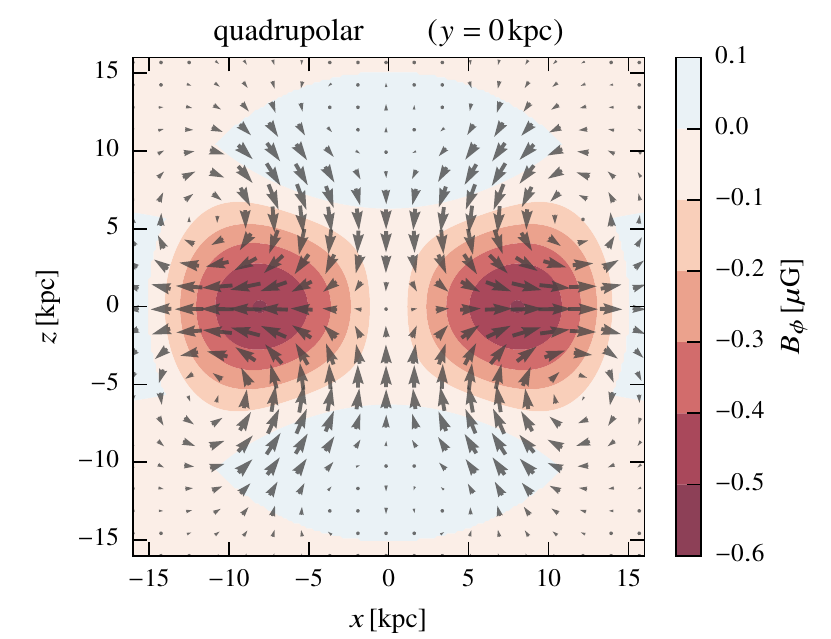}\qquad
\includegraphics[width=0.9\columnwidth]{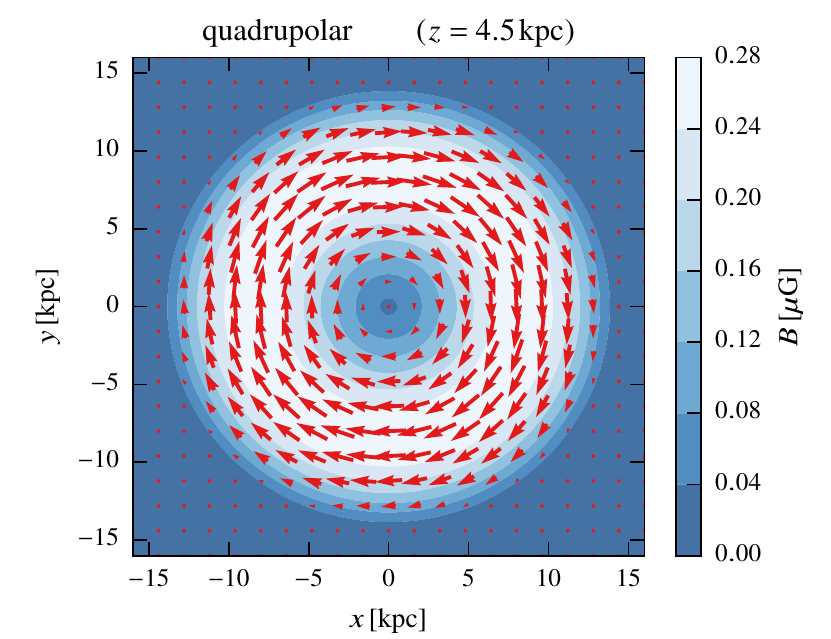}
 \caption{Examples of magnetic field configurations in the halo
 in the vertical (left) and horizontal (right) planes. The top row shows a
 magnetic structure anti-symmetric with respect to the galactic mid-plane (the dipolar
 symmetry), while the bottom row shows the symmetric magnetic field (quadrupolar symmetry).
The strength of the azimuthal magnetic field is shown with colour in the left-hand column
whilst the total field strength is colour-coded in the right-hand column. 
Arrows represent the direction and strength of the magnetic field projected onto the figure plane.
          }
 \label{fig:halo_example}
\end{figure*}

\subsection{The perturbation solution}\label{TPS}

Equation~\eqref{eq:MFDhalo} can be conveniently written 
as
\begin{equation}
    \frac{\partial \vec{B}}{\partial t} =
        \What\vec{B}+\nabla^2\vec{B}\,,
        \label{eq:MFDhalo_operator}
\end{equation}
where the perturbation operator $\What$ corresponding to the $\alpha^{2}\omega$-dynamo
is given by
\begin{equation}
    \What\vec{B}=\Rah\nabla\times({\alpha}\vec{B})+
    \Roh\nabla\times(\vec{V}\times\vec{B})\,.
    \label{eq:What_alpha2omega}
\end{equation}
As discussed in Appendix~\ref{ap:alphaOmega}, \galmag has also an option to use 
the $\alpha\omega$-dynamo operator but this approximation may be questionable in 
the case of the halo.

We substitute Eq.~\eqref{expan} into Eq.~\eqref{eq:MFDhalo_operator},
take the scalar product of the result with $\vec{B}_i$ and integrate over
the whole space. As a result, we obtain a homogeneous system of algebraic equations for 
the expansion coefficients $a_i$ of the form
\begin{equation}
    a_j(\gamma_j-\Gamma)+\sum_{i=1}^N a_iW_{ij}=0\,,
        \quad j=1,2,\ldots,N\,,\label{eq:coefficient}
\end{equation}
where
\begin{equation}\label{Wij}
    W_{ij}=\int_V\vec{B}_i\cdot\What\vec{B}_j\,{\dd}^3\vec{r}
\end{equation}
are the matrix elements of the perturbation operator, with integration performed over the 
whole space. Since the operator $\What$ transforms a poloidal field into a toroidal 
one and vice versa (and the two are orthogonal), it follows that $W_{ii}=0$ and each 
non-vanishing matrix element involves {at least} one toroidal and one poloidal free-decay
eigenfunction. Because the toroidal eigenfunctions vanish at $r>r\h$, the integrals are in 
fact restricted to the interior of the halo. The solvability condition of the system
of equations for $a_i$, the vanishing of its determinant, yields the growth rate 
$\Gamma$. Once the matrix elements have been computed and the system
\eqref{eq:coefficient} has been solved for $a_i$, Eq.~\eqref{expan} 
 yields the solution of the dynamo equation. One of the coefficients $a_i$
remains arbitrary because the dynamo equation is linear in magnetic field
and hence its solution is determined up to an arbitrary factor. This
freedom is used to fix the magnetic field strength at any desired value.

\begin{figure}
 \centering
 \includegraphics[width=0.9\columnwidth]{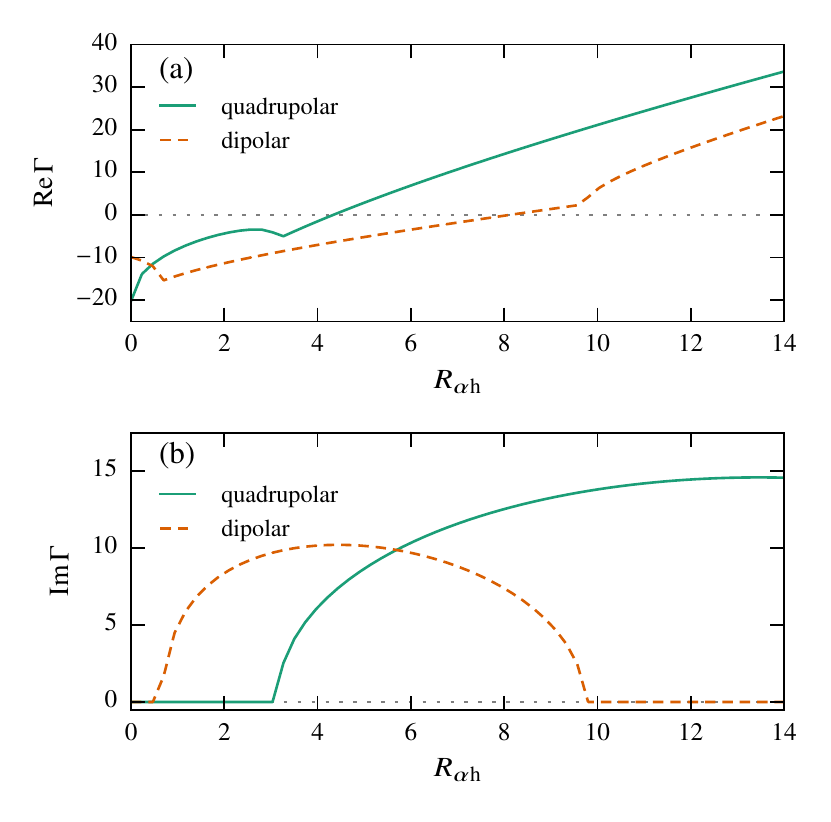}
 \includegraphics[width=0.9\columnwidth]{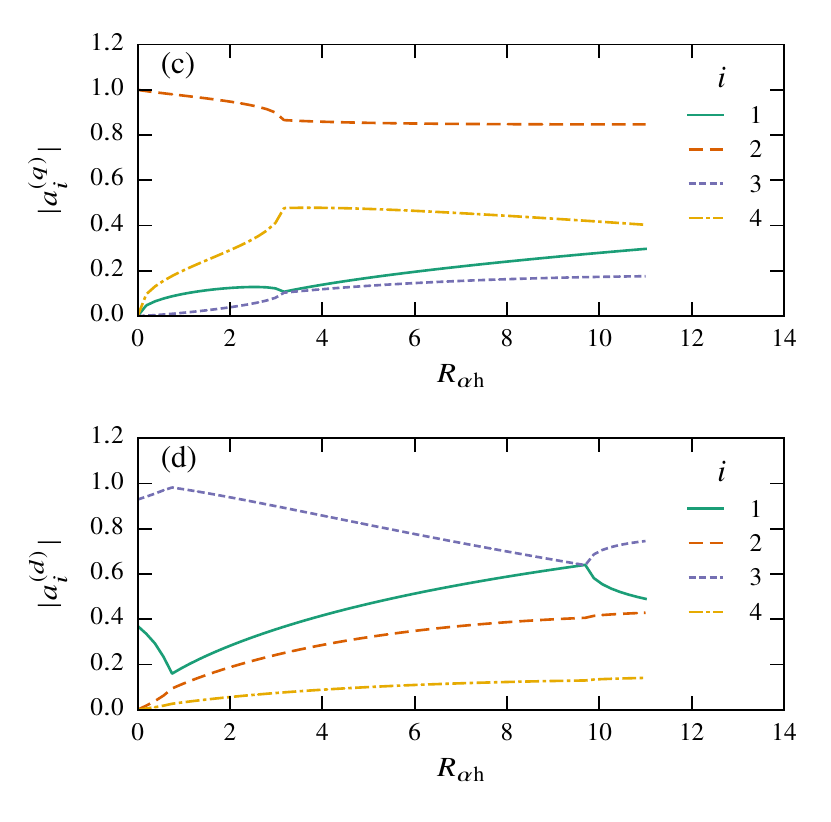}
 \caption{(a) Growth rates $\re\Gamma$ and (b) oscillation frequencies
 $\im\Gamma$ of the symmetric (solid) and anti-symmetric (dashed)
 fastest-growing dynamo eigenmodes in a spherical halo as a function of
 $\Rah$, with all the remaining parameters fixed at their fiducial values 
 shown in Table~\ref{parameters}. (c) and (d):  the magnitudes of the expansion
 coefficients $a_i$ in Eq.~\eqref{expan} for the dipolar and quadrupolar modes,
 respectively. We note that each sharp bend in
 $\re\Gamma$ that occurs as $\Rah$ changes is connected with an
 intersection of two curves $a_i$ with distinct values of $i$.
 }
 \label{fig:halo_growth}
\end{figure}

\subsection{\label{PGH}Parameters of galactic haloes}
Velocity fields in galactic haloes are poorly known. Random velocities are likely
to increase with altitude, and \ion{H}{i} observations of \citet{KWMHB98} 
\citep[see also][]{Kalberla2009} suggest a three-dimensional velocity dispersion of 
about $100\kms$, close
to the sound speed at a temperature $10^6\,\text{K}$. The scale of these motions is
uncertain. The size of supernova remnants above the galactic disc is expected to be of 
order $0.3\kpc$ \citep{MKO77}. The size of the hot gas bubbles rising from the disc
and the scale of the Parker instability are of order $0.5\text{--}1\kpc$ 
\citep[e.g.][]{RSSBF16}. Adopting the random speed and scale as $v=100\kms$ and
$l=0.5\kpc$, the turbulent diffusivity is estimated by 
$\etat\h\simeq \tfrac13lv=5\times10^{27}\cm^2\s^{-1}$. The corresponding magnetic
diffusion time across the halo radius is $r\h^2/\etat\h\simeq1.4\times10^{10}\yr$.

The knowledge of the variation of the rotation speed with position within galactic 
haloes is rather rudimentary. Both the rotational speed and its radial gradient decrease
in spiral galaxies with distance from the mid-plane, with a typical vertical gradient of 
order $\partial V/\partial z=-(15\text{--}25)\kms\kpc^{-1}$ within a few kiloparsecs from the mid-plane
\citep{ZRW15}. In our fiducial model, the halo is assumed to have a rotation curve of the 
form
\begin{equation}
 \vec{V}(\vec{r}) = V\h f(r,\theta) \, \hat{\vec{\phi}}\,
    \label{eq:Vhalo}
\end{equation}
(expressed in terms of dimensional variables), with $\hat{\vec{\phi}}$ the unit azimuthal vector and
\begin{equation}\label{eq:Vhalo2}
  f(r,\theta) = \frac{1-\exp\left(
    {-\red{\rcyl}/s_\mathrm{v}}\right)}{
     1-\exp\left(-r\h/s_\mathrm{v}\right)}\,, \quad\red{\text{with}\,s=r\sin\theta\,}   ,
\end{equation}
where the turnover radius is chosen to be $s_\mathrm{v}=3\kpc$, the typical 
value found in observations and simulations of MW-type galaxies
\citep{Reyes2011,Schaller2015}. 
For simplicity, the rotation curve of Eqs.~\eqref{eq:Vhalo} and~\eqref{eq:Vhalo2} 
has no $z$-dependence but it can easily be introduced. The role of the variation of
$\Omega$ with $z$ is to produce $B_\phi$ from $B_z$, arguably a process somewhat
less important than the stretching of the radial magnetic field in the azimuthal 
direction at a rate $S=\rcyl\partial\Omega/\partial\rcyl$.

\begin{figure}
 \centering
 \includegraphics[width=\columnwidth]{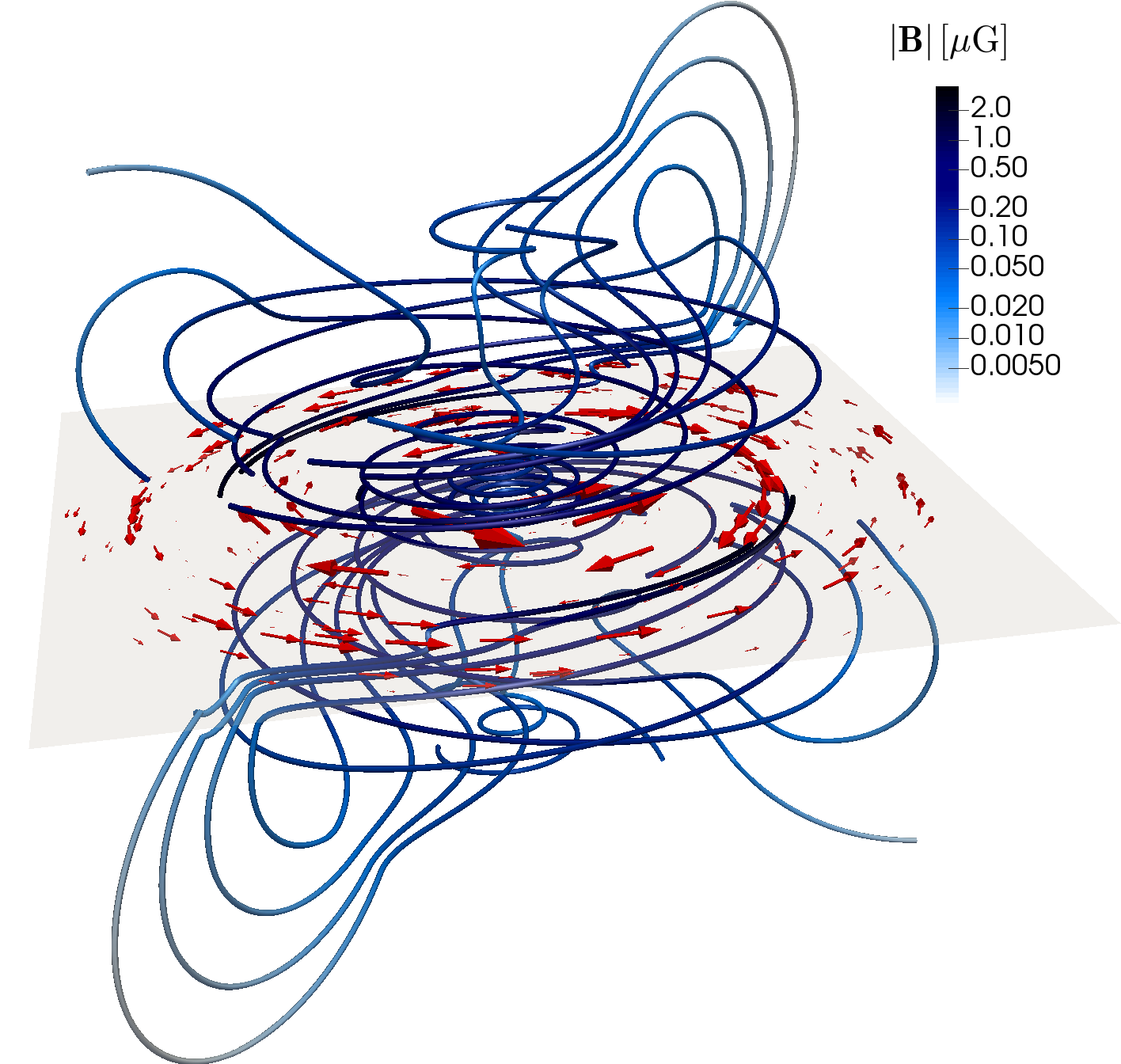}
 \caption{Three-dimensional rendering of a symmetric (quadrupolar) halo
    field combined  with a quadrupolar disc field with two reversals
    at $\rcyl=7\kpc$ and $12\kpc$. The domain is a $(17\kpc)^3$ box.
    The field lines were seeded uniformly along a diagonal through the box.
    The arrows show the magnetic field at points randomly sampled within the slice
    of a thickness $2.5\kpc$ around the galactic mid-plane (which is indicated by
    the semi-transparent surface) and are scaled according to the magnitude of
    the magnetic field.}
 \label{fig:renderings}
\end{figure}
\begin{figure}
 \centering
 \includegraphics[width=0.8\columnwidth]{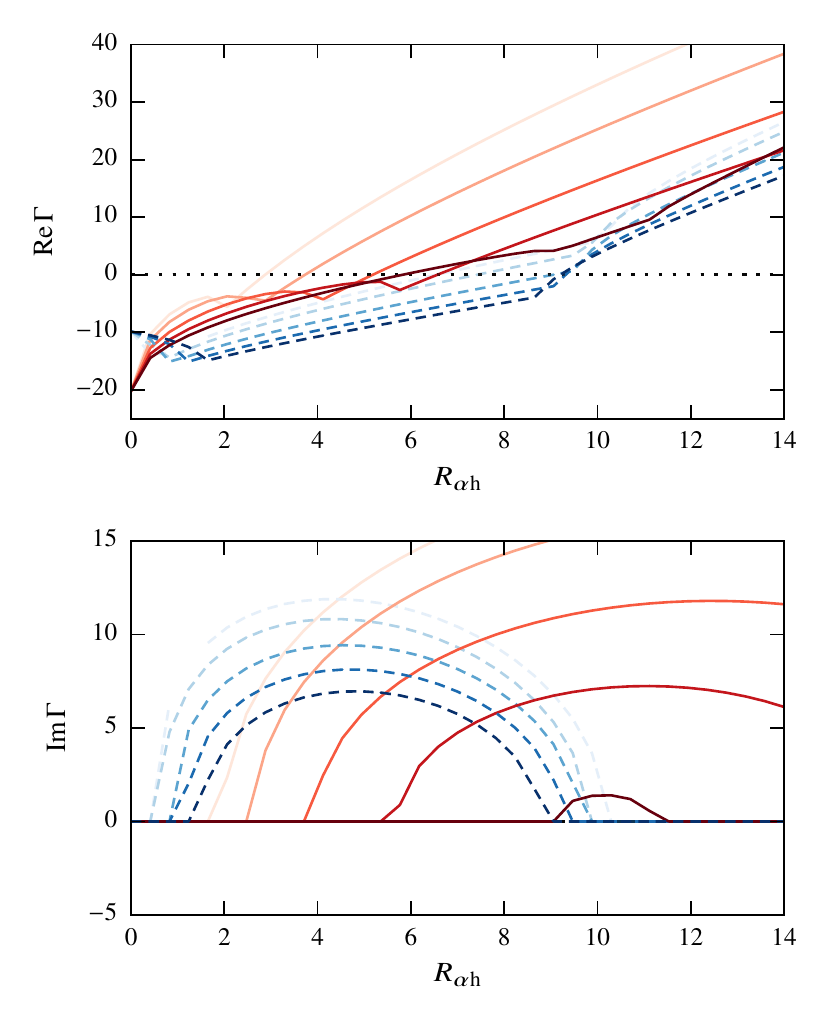}
 \caption{The effect of the form of the rotation curve on the dynamo growth rate
 $\re\Gamma$ and oscillation frequency $\im\Gamma$ of the fastest-growing
 magnetic field in the spherical halo. The shade of each curve corresponds to
 the value of the turnover radius of the rotation curve, $\rcyl_\text{v}$, with the
 lightest shade for $\rcyl_\text{v}=0.5\kpc$ and the darkest for $\rcyl_\text{v}=7.5\kpc$
 with the increment of $1.75\kpc$ (the fiducial value is $\rcyl_\text{v}=3\kpc$).
 Dashed curves are for the anti-symmetric eigenmodes and solid curves show
 symmetric solutions.}
 \label{fig:halo_growth_sv}
\end{figure}

We adopt a simple form for the $\alpha$-coefficient often used in spherical mean-field
dynamo models; in dimensional variables,
\begin{equation}
\alpha(\vec{r}) = \alpha\h \cos\theta\,,
\end{equation}
implying the largest absolute value of $\alpha$ near the poles whilst $\alpha$ also vanishes at the equator, 
reflecting the fact that the mean helicity of the random flows is produced by the 
Coriolis force.

We consider an axially symmetric magnetic field in the halo and assume that the
dynamo operates within a region of $r\h = 15\kpc$ in radius. We take
$V\h=220\kms$ (similar to that in the disc). With the turbulent magnetic diffusivity
$\etat\h=5\times10^{27}\cm^2\s^{-1}$, this leads to $ \Roh\simeq200$.

Estimating $\Rah$ in the halo is more difficult given the uncertainty 
of the random flow parameters. The standard estimate 
of Eq.~\eqref{eq:an} yields $\alpha\h\simeq l^2\Omega/h\simeq1\kms$ and 
${\Rah=\alpha\h r\h/\etat\h\simeq1}$
for $\Omega=26\kms\kpc^{-1}$ and 
$h=3\kpc$, the gas density scale height in the halo. As the fiducial value for 
$\Rah$, we select its marginal value corresponding to the vanishing dynamo growth 
rate (see Section~{\ref{MSH}} for details): $\Rah^\text{(q)}=4.3$ for
symmetric solutions and $\Rah^\text{(d)}=8.1$ for anti-symmetric ones. 
The symmetric mode is preferred to the anti-symmetric one only slightly, 
$\Rah^\text{(q)}/\Rah^\text{(d)}\simeq0.5$ for the marginal values. The 
similarity of the marginal values of $\Rah$ for the dipolar and quadrupolar 
magnetic structures in the halo reflects the fact that, unlike the disc dynamo, 
spherical dynamos usually do not exhibit a strong preference for either symmetry.

\subsection{Basic magnetic structures}\label{MSH}
Figure~\ref{fig:halo_example} shows two examples of magnetic structures in the halo 
that are marginally stable with respect to the mean-field dynamo action,
$\partial\vec{B}/\partial t=\vec{0}$, one symmetric with respect to the equator and 
the other anti-symmetric:
\begin{align*}
  \vec{B}_\text{h}^\text{(d)} &\approx -0.48 \vec{B}_1^\text{(d)}
                        -0.38 \vec{B}_2^\text{(d)}
                        -0.70 \vec{B}_3^\text{(d)}
                        -0.12 \vec{B}_4^\text{(d)}\,,\\
  \vec{B}_\text{h}^\text{(q)} &\approx 0.14 \vec{B}_1^\text{(q)}
                        +0.86 \vec{B}_2^\text{(q)}
                        +0.10 \vec{B}_3^\text{(q)}
                        -0.41 \vec{B}_4^\text{(q)}\,.
\end{align*}
The eigenfunctions are normalised to have 
$B_\text{h}^\text{(d)}=-0.5\muG$ and $B_\text{h}^\text{(q)}=-0.01\muG$ at 
$(\rcyl,z)=(8.5,0.02)\kpc$ in the anti-symmetric and symmetric cases, respectively, 
so that they have similar maximum magnetic field strengths.

The poloidal magnetic lines (the left-hand panels) have the so-called X shape detected
in the halos of some galaxies, especially pronounced in the quadrupolar
structure. This is a generic field structure typical of any divergence-free vector field that can be enhanced further by a large-scale velocity
shear of the galactic wind or fountain. Unlike the symmetric eigenfunction, the anti-symmetric
one has a maximum away from the equator. The position of the maxima depends on the spatial
forms of $\alpha(\vec{r})$ and $\vec{V}(\vec{r})$; in our case, the rotation speed is
independent of $z$, and $\alpha(\vec{r})$ alone controls this feature.

It is not clear which of the two symmetries may dominate in galactic haloes: this depends on
the strength of the magnetic coupling between the disc and the halo and between the two
hemispheres of the halo. If the disc-halo coupling is strong or the disc disrupts magnetic
connection between the two halo hemispheres, the quadrupolar disc field could enforce a
symmetric field structure in the halo. Halo fields of mixed parity are also a possibility,
but their modelling requires non-linear dynamo solutions rather than superpositions of linear
eigenmodes that we use here. The 
strength of the disc-halo magnetic connection depends on the ratio of the turbulent magnetic 
diffusivities in the two regions: the larger the value of  $\etat\h/\etat\di$, the weaker the coupling. 
Existing models of the mean-field dynamo action in galactic disc-halo systems only considered 
the range $\etat\h/\etat\di\leq30$.

\begin{figure}
 \centering
 \includegraphics[width=\columnwidth]{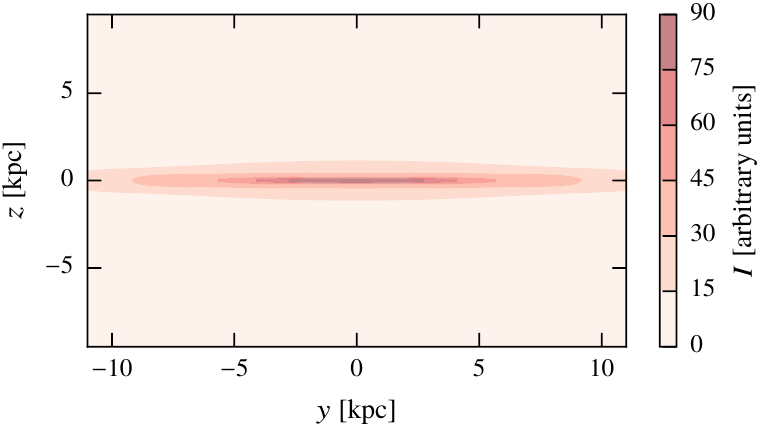}
 \includegraphics[width=\columnwidth]{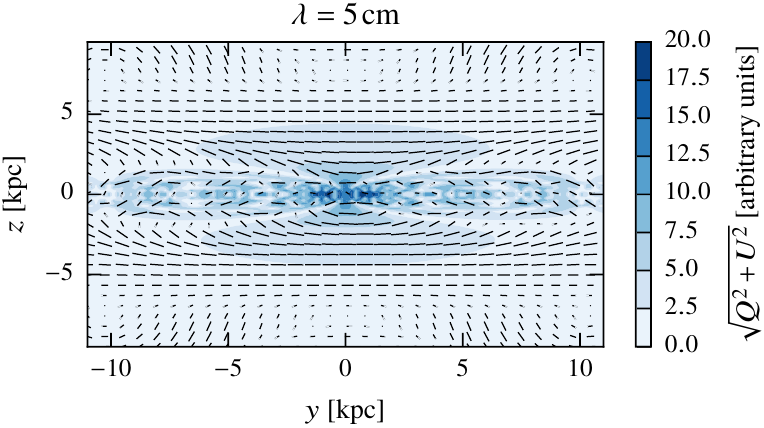}
 \includegraphics[width=\columnwidth]{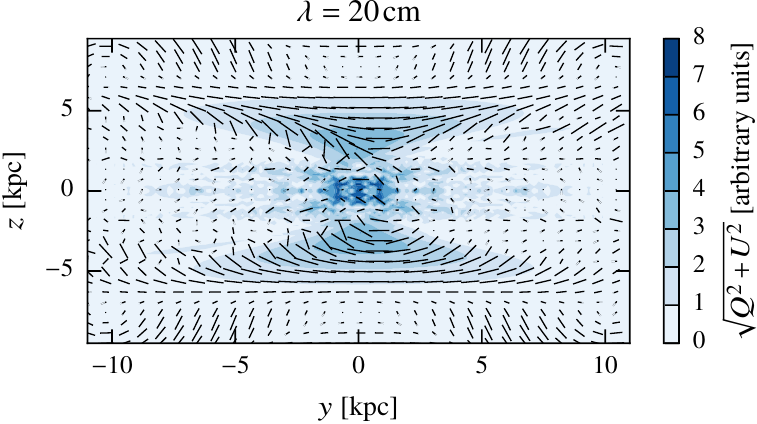}
 \caption{Synchrotron emission produced by the magnetic configuration shown 
 in Fig.~\ref{fig:renderings} with the disc seen edge-on (see the text for other assumptions). Top
 panel: the total intensity (Stokes parameter $I$). Middle and bottom panels:
 the polarised intensity at $\lambda=5$ and $\lambda=20\cm$, respectively.
 The dashes are perpendicular to the polarisation angle and their lengths 
 are proportional to the fractional polarisation. No correction for random
 magnetic fields has been made in the fractional polarisation.
          }
 \label{fig:synchrotron}
\end{figure}

The top two panels of Fig.~\ref{fig:halo_growth} illustrate how the growth rates 
and oscillation frequencies of the
symmetric and anti-symmetric modes depend on $\Rah$ when all other parameters are fixed
to the fiducial values of Table~\ref{parameters}. These solutions involve
the first four symmetric or anti-symmetric free-decay modes
with expansion coefficients shown in the lower half of Fig.~\ref{fig:halo_growth}. 
As shown in Fig.~\ref{fig:halo_growth}b  , both the symmetric and anti-symmetric 
eigenmodes are typically oscillatory ($\im\Gamma\neq0$). The fact that $\im\Gamma=0$ for the anti-symmetric 
mode at $\Rah\gtrsim10$ appears to be an artefact of including only a small
number of the free-decay modes into the perturbation series. The series \eqref{expan}
converges rather slowly \citep{RadWie89,RWBMT90} and adding a few more terms does not always 
improve the accuracy \citep{SNEL08}. Therefore, the model for the halo magnetic field
that involves only a modest number of modes can reproduce only relatively simple
magnetic configurations (and yet quite non-trivial -- see Fig.~\ref{fig:renderings}). This does 
not appear to be a serious problem, though, since the scale of the mean magnetic field in 
galactic haloes is unlikely to be smaller than a few kiloparsecs.

The dependencies of the growth rate and oscillation frequency of the magnetic field on 
the turnover radius of the rotation curve $\rcyl_v$ are shown in
Fig.~\ref{fig:halo_growth_sv}. Larger values of $\rcyl_v$ correspond to weaker differential 
rotation and, therefore, lower growth rates and oscillation frequencies.  

\section{Discussion}\label{sec:discuss}

Magnetic fields obtained above for the disc and halo are combined by a simple superposition 
with arbitrary weights. For illustration, we show in Fig.~\ref{fig:renderings} a magnetic 
structure that has two radial field reversals in the disc (Model~B of
Fig.~\ref{fig:disc_field}) and a symmetric (quadrupolar) field in the halo shown in 
the bottom row of Fig.~\ref{fig:halo_example}. The complexity of the resulting magnetic
structure clearly illustrates the possibilities of the model. The most important 
limitations of the model in its current form is that it is axially symmetric and does not
include galactic outflows. Both can be addressed rather straightforwardly within the
framework of this approach. In the following we discuss some applications and  
extensions of our approach.

\subsection{Synthetic radio maps}\label{sec:radio}

The model can be used to interpret observations of synchrotron emission and Faraday rotation
as soon as the distributions of cosmic ray and thermal electrons have been specified.  
The total and polarised synchrotron intensities, polarisation angle and Faraday rotation
measure can be derived as described in Appendix~\ref{ap:polarisation}.
As a simple illustrative example, we assume a uniform distribution for the 
cosmic-ray electrons, $n_\gamma=\const$ in terms of the number density, and
adopt exponential profiles for the number density of thermal electrons,
\begin{equation}
n_\e(\rcyl, z, \phi) = n_0 
    \exp\left[-\frac{z}{h(\rcyl)}-\frac{\rcyl}{\rcyl_\text{e}}\right]\,,
    \label{eq:ne}
\end{equation}
where $h(\rcyl)$ is given by Eq.~\eqref{eq:scaleheight} and $\rcyl_\e=3\kpc$, is the
scale radius of the disc (chosen to be similar to the case of the Milky Way,
\citealt{BinneyTremaine2008}). Any other model \citep[e.g.][]{CL02} could be used
instead but we prefer to avoid exaggerating the amount of detail in the magnetic field
model that may arise from details such as spiral arms in more complicated models for
$n_\e$.

In the top panel Fig.~\ref{fig:synchrotron}, we show the synchrotron emission in a 
galaxy seen edge-on with the magnetic field of Fig.~\ref{fig:renderings}. 
The other two panels show the polarised emission at two wavelengths, $\lambda=5\cm$ and
$\lambda=20\cm$, as in C- and L-bands of the VLA used, for example, in the CHANG-ES survey
\citep{Irwin2012}. At $5\cm$, most of the polarisation signal
is dominated by the disc component and localised around the mid-plane of the disc.
At longer wavelengths, most of the emission from the galactic plane is depolarised and 
two conical lobes of about $5\kpc$ in height are prominent in the halo, similar to 
the so-called X-shaped structures observed in edge-on galaxies \citep{Wiegert2015}.

\subsection{Evolution of galactic magnetic fields}\label{sec:Bevol}

Another possible application is a simple, approximate model for the evolution of
a large-scale magnetic field in a galaxy. In this application, an initial (seed)
magnetic field has to be prescribed and then it can be evolved using the
growth rates of the magnetic modes derived above. Suitable initial conditions 
must then be selected. The simplest approach is to assume
that all the modes are equally represented in the initial state, 
that is, $C_n$ independent of $n$ and chosen to obtain an initial large-scale magnetic field of
any given strength. A physically better motivated initial magnetic field represents a
random field produced by the fluctuation dynamo in a young galaxy or protogalaxy \citep{PSS93}. Because 
of the finite size of the dynamo region, the projection of such a random field onto the
dynamo eigenmodes does not vanish. \citet[Sect.~VII.14 in][]{RSS88} estimate the
corresponding initial dimensional magnitudes of the radial disc modes as
\begin{equation}
C_n^{(0)}=\frac{b}{N_n^{1/2}}\,\frac{l}{\delta\rcyl_n}\,,
\end{equation}
where $b$ is the root-mean square strength of the random magnetic field $\vec{b}$, 
$N_n$ is the number of the correlation cells of $\vec{b}$ within a cylindrical annulus 
of an axial extent $2h$, radius $\rcyl$ and width  $\delta\rcyl_n$, with 
$\delta\rcyl_n$ the radial scale of $Q_n(\rcyl)$, and $l$ is the scale of $\vec{b}$. 
With $\delta\rcyl\simeq\rcyl_\text{disc}/n$, $N\simeq h\rcyl\rcyl_\text{disc}/(nl^3)$, $l=100\p$, $\rcyl_\text{disc}=20\kpc$, 
we have
\begin{equation}\label{Cn}
C_n^{(0)}\simeq n^{3/2} \left(\frac{\rcyl}{\rcyl_\text{disc}}\right)^{-1/2} 
	\left(\frac{l^5}{h\rcyl_\text{disc}^4}\right)^{1/2}
		\simeq 10^{-5} n^{3/2}b\left(\frac{\rcyl}{\rcyl_\text{disc}}\right)^{-1/2}.
\end{equation}
Thus, the seed for the large-scale dynamo due to the small-scale magnetic field  
favours higher-order modes being proportional to $n^{3/2}$, because they have smaller 
scale, and decreases with radius as $\rcyl^{-1/2}$. A plausible estimate is 
$b\simeq5\muG$ by analogy with observational estimates for nearby spiral galaxies.
Otherwise, if a dependence on the interstellar gas parameters is required,
a suitable estimate is
\begin{equation}
b\simeq (4\pi\rho)^{1/2}v\,,
\end{equation}
where $\rho$ is the gas density in the diffuse warm interstellar gas and $v$ is the
turbulent speed. The standard estimates of the latter are
$\rho\simeq1,7\times10^{-24}\g\cm^{-3}$ corresponding to the number density of
$1\cm^{-3}$ and $v\simeq10\kms$. This yields $b\simeq3\muG$.

\subsection{Extensions of the model}\label{sec:EM}
There are several directions in which the model can be extended. Perhaps most important
is to include non-axisymmetric magnetic fields. This is straightforward to implement. 
In the disc, the local equations of Section~\ref{sec:disk_local} and their solutions
remain unchanged but the radial part of the eigenfunction $Q_n$ of
Section~\ref{sec:disk_radial} becomes a function of both radius and azimuth. Solutions 
for $Q_m(\rcyl,\phi)$ were obtained in largely the same manner as above by
\citet{BSRS87} \citet{KSRS89} and \citet{BPSS97}, and are reviewed by 
\citet[Sect.~VII.8 in][]{RSS88} and \citet{KRSS90}.
Introducing non-axisymmetric magnetic fields in the halo would only require that 
non-axisymmetric free-decay modes are included into the perturbation solution. This is
straightforward to do and does not require any significant modification of the formalism of
Section~\ref{TPS}.

Another physically important generalisation is the inclusion of galactic outflows and 
accretion flows, that is, large-scale poloidal velocity fields $\vec{U}$. The additional
velocity components appear in the perturbation operators. Within the disc, $U_z$ 
enters the local equations \eqref{eq:locals}--\eqref{eq:MFDc_azimuthal_alt} whereas
$U_\rcyl$ is included in the radial equation~\eqref{eq:Qs}. The modified solutions
are discussed by \citet{BvRDBS01} and \citet{MoShSo00}, respectively. In the halo,
the poloidal velocity just enters the perturbation operator \eqref{eq:What_alpha2omega}
without affecting the procedure of perturbation analysis.

The solutions used in the model are kinematic (linear in magnetic field) as they are
derived for $\vec{V}$, $\alpha$ and $\etat$ independent of $\vec{B}$. The linear 
nature of the solution is not restrictive in the present context since its aim is to 
provide a convenient functional basis to parametrise a desired magnetic configuration. 
On the other hand, \citet{CSSS14} show that a wide class of non-linear solutions are well
approximated by the marginally stable eigenfunction (i.e., that obtained for
$\partial\vec{B}/\partial t=0$). Nonlinear dynamo effects, leading to solutions sensitive
to the gas density and other relevant parameters, can be introduced in the radial thin-disc
equation~\eqref{eq:Qs} as discussed by \citet{PSS93}, via a non-linear modification
(quenching) of the local growth rate which becomes a function of $Q$:
\[
\gamma(\rcyl,Q) = \gamma(\rcyl)\left[ 1- Q^2/B_0^2(\rcyl)\right]\,,
\]
where $\gamma(\rcyl)$ is the kinematic local growth rate obtained as discussed in 
Section~\ref{sec:disk_local}. Since the time scale of magnetic field evolution in the
halo is comparable to $10^{10}\yr$ (Section~\ref{sec:SDS}), non-linear dynamo
effects are likely to be less important in galactic haloes.

\subsection{The \textsc{galmag} software package}\label{galmag}

The model presented in this paper has been implemented as the Python software package
\galmag\footnote{\url{https://github.com/luizfelippesr/galmag}} \citep{GALMAG_zenodo},
which is publicly available under the GNU General Public License v3.
Further details can be found in the on-line code 
documentation\footnote{\url{http://galmag.readthedocs.io/}}, which includes a tutorial.

Since \galmag uses Python objects of the \textsc{d2o} package \citep{Steininger2016} 
instead of regular \textsc{numpy} arrays,
when it is invoked using MPI, all the array operations are automatically performed
in parallel. As a stand-alone package, it can synthesise three-dimensional magnetic 
field structures from a provided set of expansion coefficients and compute synthetic maps
of the Stokes parameters of synchrotron emission and Faraday rotation. 

A more flexible use of \textsc{galmag} is to employ it in modular magnetic field 
optimisation frameworks like the IMAGINE pipeline \citep{Steininger2018, Steininger2018ASCL}, 
where it serves as a magnetic field generator and is interfaced to multi-purpose observable 
generators, such as the \hammurabi code\footnote{\url{https://sourceforge.net/p/hammurabicode/wiki/Home/}}
\citep{waelkens:2009}. This allows us to not only compute maps of observables from any point of view 
and thus to compare with observations, but also provides sophisticated sampling techniques to optimise 
the \textsc{galmag} parameters.   

\section{Conclusions}\label{sec:final}

We have presented an approach to develop parametrised models for large scale magnetic 
fields of the Milky Way and other disc galaxies based on fundamental equations 
of magnetic field generation and evolution. Implemented in the software package \textsc{galmag}, 
it is designed to be used in interpretations of observations of Faraday rotation, 
synchrotron and dust emission, and other observational tracers. 
In this paper, we have presented the basic formalism of the approach, and demonstrated its 
capabilities in illustrative examples.

The model is based on the expansion of the large-scale magnetic field 
over a basis of eigenfunctions of the mean-field dynamo equation \eqref{MFD}, and the standard 
induction equation is its special case obtained for $\alpha=0$. As 
long as the functional basis is complete, any magnetic structure, whether or not produced 
by the dynamo, can be represented as a superposition of the eigenfunctions. Therefore,
an alternative use of the magnetic field model is to represent any magnetic configuration 
of interest in terms of a relatively small number of parameters. The resulting magnetic 
field is physically realisable, being a solution of the induction equation or its
modification with $\alpha\neq0$, as desired. Furthermore, the fact that the model 
parameters have clear physical meaning would help to refine it so as to satisfy any
additional constraints. Magnetic fields of the model are obtained in the form of
series expansions, and the series can  be truncated to achieve the desired amount of detail 
in the resulting solution.

The novelty and strength of our approach lies in advancing both 
the flexibility and physical plausibility of GMF models. It will be useful in Bayesian 
optimization machines, which seek among many reasonable morphological approaches to the GMF 
structure the one that gives the best results both in terms of physical plausibility and the 
explanation of existing data.

\begin{acknowledgements}
We have benefited from fruitful discussions with the members of the 
\href{http://www.issibern.ch/teams/bayesianmodel/}{ISSI International Team 323},
\textit{Bayesian modeling of the Galactic magnetic field constrained by space- and
ground-based radio-millimetre
and ultra-high energy cosmic ray
data\footnote{\url{http://www.issibern.ch/teams/bayesianmodel/}}},
and the \href{https://www.astro.ru.nl/imagine/}{IMAGINE 
Consortium}\footnote{\url{https://www.astro.ru.nl/imagine/}}.
Special thanks are due to Torsten En{\ss}lin, Marijke Haverkorn,
Jens Jasche and Andrew Fletcher for their useful comments, and Theo Steininger
for help with the software development and the \textsc{d2o} Python package 
AS, LFSR and JPR acknowledge financial support and hospitality
of the International Space Science Institute (ISSI) in Bern, Switzerland.
AS and LFSR are supported by STFC (ST/N000900/1, Project 2), and AS and PB are supported by the 
Leverhulme Trust (RPG-2014-427). This research has made use of NASA's Astrophysics Data System.
\end{acknowledgements}
{
\bibliographystyle{aa} 
\bibliography{GMF}
}

\begin{appendix}
\section{Spherical free-decay modes}\label{ap:free_decay}

For axisymmetric free-decay modes, a solution to Eq.~\eqref{eq:free_decay} is obtained
in terms of scalar potentials, as discussed by \citet{KR80} and \citet{M78}.
For the reader's convenience, we present an outline of the solution. 
Any magnetic field $\vec{B}$ can be represented as the sum of a poloidal field 
$\nabla\times\vec{A}_\text{P}$, where $\vec{A}_\text{P}$ is its vector potential, and a 
toroidal field $\vec{B}_\text{T}$:
\begin{equation}\label{eq:pplust}
\vec{B}=\nabla\times\vec{A}_\text{P}+\vec{B}_\text{T}\,,
\end{equation}
and $\nabla\cdot\vec{B}=0$ provided 
\begin{equation}
  \vec{A}_\text{P}=-{\vec{r}}{\times}{\nabla}S\,, \qquad
  \vec{B}_\text{T}=-{\vec{r}}{\times}{\nabla}T\,,
  \label{eq:potentials}
\end{equation}
where $\vec{r}$ is the position vector normalised such that $r=1$ is the
halo surface ($r=r\h$ in dimensional variables)
and $S$ and $T$ are known as the scalar potentials. In terms of the scalar 
potentials and assuming axial symmetry, Eq.~\eqref{eq:free_decay} reduces 
in spherical coordinates $(r,\theta,\phi)$ to 
\begin{align}
\frac{1}{r^{2}}\frac{\partial}{{\partial}r}\left({r^{2}}
\frac{{\partial}S}{{\partial}r} \right) +
\frac{1}{r^{2}\sin{\theta}}\frac{\partial}{{\partial}\theta}
\left({{\sin{\theta}}}\frac{{\partial}S}{{\partial}\theta} \right)=\gamma S\,,
\label{eq:potentialequationsa} \\
\frac{1}{r^{2}}\frac{\partial}{{\partial}r}\left({r^{2}}
\frac{{\partial}T}{{\partial}r} \right) +
\frac{1}{r^{2}\sin{\theta}}\frac{\partial}{{\partial}\theta}
\left({{\sin{\theta}}}\frac{{\partial}T}{{\partial}\theta} \right)=\gamma T\,,
\label{eq:potentialequationsb}
\end{align}
for $r<1$ and,
\begin{equation}
\nabla^2S=0\,,\qquad T=0\qquad \text{for}\ r>1\,,
\end{equation}
with the vacuum boundary conditions
\begin{equation}
[S]=[\partial S/\partial r]=[T]=0 \ \text{at}\ r=1\,,
\label{eq:potenitalbcs}
\end{equation}
where $[X]$ denotes the jump of $X$, and $[X]=0$ means continuity.
We also require both potentials to be finite at $r=0$.

The potentials satisfy identical equations at $r<1$, so consider this
equation for $G$ equal to either $S$ or $T$,
\begin{equation}
\frac{1}{r^{2}}\frac{\partial}{{\partial}r}\left({r^{2}}
\frac{{\partial}G}{{\partial}r} \right) +
\frac{1}{r^{2}\sin{\theta}}\frac{\partial}{{\partial}\theta}
\left({{\sin{\theta}}}\frac{{\partial}G}{{\partial}\theta} \right)
-{\gamma}G=0\,.
\label{eq:generalpotential}
\end{equation}
Using separation of variables, $G(r,\theta)=R(r)\Theta(\theta)$, Bessel's equation is 
obtained in $r$ and Legendre's equation in $\theta$, with the separation constant $n(n+1)$ 
($n=1,2,3,\ldots)$:
\begin{align}
r^{2}\oderiv{^2R}{r^2}+2r\oderiv{R}{r}-[{\gamma}r^{2}+n(n+1)]R=0\,,
\label{eq:Bessel} \\
 \oderiv{}{\theta}\left(\sin\theta\oderiv{\Theta}{\theta}\right)+n(n+1)\Theta\sin\theta=0\,.
\label{eq:Legendre}
\end{align}
In terms of $x=\sqrt{-\gamma}r$ and $Q(x)=x^{1/2}R(x)$ in Eq.~(\ref{eq:Bessel}) 
and $x=\cos\theta$ in Eq.~(\ref{eq:Legendre}), we have
\begin{align}
x^2\oderiv{^2Q}{x^{2}}+x\oderiv{Q}{x}+\left[x^2-\left(n+\tfrac{1}{2}\right)^2\right]Q=0\,, \label{eq:Bessel_1} \\
\oderiv{}{x}\left[(1-x^2){\oderiv{\Theta}{x}} \right]+n(n+1)\Theta=0.
\label{eq:Legendre_1}
\end{align}
Non-singular solutions of \eqref{eq:potentialequationsa} and 
\eqref{eq:potentialequationsb} then follow as
\begin{align}
T&=\sum\limits_{n=1}^{\infty}\sum_{l=1}^{\infty} c_{nl}
			T_{nl}(r)P_{n}(\cos\theta)\,, \\
S&=r_\text{h}\sum_{n=1}^{\infty}\sum_{l=1}^{\infty} c_{nl}
			S_{nl}(r)P_{n}(\cos\theta)\,,
\label{eq:potentialsolutons}
\end{align}
where
\begin{equation}\label{eq:Tnlsnl}
T_{nl}(r)=S_{nl}(r)=\frac{1}{\xi_{nl}\sqrt{r}}
						J_{n+1/2}(\xi_{nl}r)\,,
\end{equation}
with constants $c_{nl}$,
\begin{equation}
\xi_{nl}=\sqrt{-\gamma_{nl}}\,,
\end{equation}
and $\xi_{nl}$ are solutions to \eqref{eq:Jcondition}. 
The factor $r_\text{h}$ in Eq.~(\ref{eq:potentialsolutons}) is introduced to 
ensure dimensional consistency when obtaining the magnetic field from these potentials. 
The boundary conditions \eqref{eq:potenitalbcs} reduce to
\begin{equation}
T_{nl}=0\,, \quad S_{nl}=d_n\,, \quad \deriv{S_{nl}}{r}=-(n+1)d_n\quad  \text{at } r=1\,,
\label{eq:potenitalbcs2}
\end{equation}
where $d_n$ are constants. Eliminating $d_n$, the
boundary conditions for $S_{nl}$ reduce to the recurrence relation
\begin{equation}
\deriv{S_{nl}}{r}+(n+1)S_{nl}=0\quad \text{at }r=1\,.
\label{eq:potenitalbcs3}
\end{equation}
Together with the requirement that $T_{nl}$ and $S_{nl}$ do not vanish simultaneously, this gives
\begin{equation}
J_{n-1/2}(\xi_{nl})J_{n+1/2}(\xi_{nl})=0\,,
\label{eq:Jcondition}
\end{equation}
which determines the admissible values for $\xi_{nl}$, and $\gamma_{nl}=-\xi_{nl}^2$
yields the decay rates $\gamma_{nl}$ given in Table~\ref{tab:decay_rates}.

For $l$ odd, $J_{n-1/2}(\xi_{nl})=0$ for all $n$. Hence, $T_{nl}=0$ for
$l$ odd. Conversely, when $l$ is even, $J_{n+1/2}(\xi_{nl})=0$ for all $n$. 
Hence, $S_{nl}=0$ for $l$ even. The solutions satisfying the boundary conditions 
can be written as follows:
\begin{subequations}\label{eq:potentialsolutions}
\begin{align}
T&=\sum_{{n=1}}^\infty\sum_{l \, \text{even}}
		\frac{c_{nl}}{\sqrt{r}}J_{n+1/2}(\xi_{nl}r)
        P_{n}(\cos\theta)\,,\\
S&=\sum_{{n=1}}^\infty\sum_{l \, \text{odd}}
		\frac{d_{nl}}{\sqrt{r}}J_{n+1/2}(\xi_{nl}r)
        P_{n}(\cos\theta)\,,
\end{align}        
\end{subequations}
where 
$c_{nl}$ and $d_{nl}$ are constants. Individual terms in 
the sums are arranged according to increasing magnitude of the decay rates, $|\gamma_{nl}|$.

\begin{table}
\centering
\caption{Decay rates of spherical free-decay modes, $\gamma_{nl}$.}
\label{tab:decay_rates}
\begin{tabular}{ccccc}
\hline\hline 
& $l=1$ & $l=2$ & $l=3$ & $l=4$ \\
\hline
$n=1$ & $-\pi^2$ 	& $-(4.493)^2$ & $-(2\pi)^2$ & $-(7.725)^2$  \\
$n=2$ &$-(4.493)^2$ & $-(5.763)^2$ & $-(7.725)^2$ & $-(9.095)^2$  \\
$n=3$ &$-(5.763)^2$ & $-(6.988)^2$ & $-(9.095)^2$ & $-(10.417)^2$ \\
$n=4$ &$-(6.988)^2$ & $-(8.813)^2$ & $-(10.417)^2$& $-(11.705)^2$ \\
\hline
\end{tabular}
\end{table}

The free-decay modes form two separate families based on their symmetry about the equator
$\theta=\pi/2$, the anti-symmetric (dipolar) and symmetric (quadrupolar) ones. 
The anti-symmetric modes, denoted with superscript (d), 
occur when both $n$ and $l$ are either odd or even,
whilst the symmetric modes that have superscript (q) occur otherwise  .

Explicit forms of a few of the lowest free-decay modes are given in the next section, each 
normalised to 
\begin{equation}\label{hnorm}
\int_V|\vec{B}_n|^2\,\dd^3\vec{r}=1\,,
\end{equation}
where the integral is taken over 
the sphere $r\leq1$, to form an orthonormal set. Although each eigenmode is either poloidal 
or toroidal, their superpositions \eqref{eq:potentialsolutions} necessarily contain both 
poloidal and toroidal parts: purely toroidal and purely poloidal fields cannot sustain 
Ohmic dissipation and unavoidably decay.

\subsection{Symmetric modes}\label{sec:free_symm}
The four leading quadrupolar free-decay modes are shown in the upper row of 
Fig.~\ref{fig:free_decay}. The quadrupolar mode of the slowest decay
has $(n,l)=(2,1)$ and is poloidal,
\begin{equation}
\vec{B}_1^\text{(q)}=A_1  \left(
\frac{\red{Q}_1(r)}{r}(3\cos^2\theta-1)\,,\
	-\frac{\sin\theta\,\cos\theta}{r}\oderiv{}{r}[r\red{Q}_1(r)]\,,\ 0\right)\,,
\end{equation}
where $A_{1} \approx 0.662$ and
\begin{equation}
\red{Q}_1(r)=  \begin{cases}
    r^{-1/2}J_{5/2}(q_1r)\,,	&r \leq 1\,, \\
    r^{-3}J_{5/2}(q_{1})\,,		&r > 1\,, \\
  \end{cases}
  \qquad q_1 \approx 4.493\,.
\end{equation}
The next mode, $(n,l)=(1,2)$, is toroidal and has the same eigenvalue,
\begin{equation}
\vec{B}_2^\text{(q)}=A_2\left(0,\ 0,\ \red{Q}_2(r)\,\sin\theta \right)\,,
\end{equation}
where $A_2 \approx 1.330$ and
\begin{equation}
\red{Q}_2(r)=  \begin{cases}
    r^{-1/2}J_{3/2}(q_1r)\,,	&r \leq 1\,, \\
    r^{-2}J_{3/2}(q_1r)\,,		&r > 1\,.
  \end{cases} 
\end{equation}

The modes $\vec{B}_3^\text{(q)}$ and $\vec{B}_4^\text{(q)}$, poloidal and toroidal 
respectively, also form a doublet with the common eigenvalue and
correspond to $(n,l)=(4,1)$ and $(n,l)=(3,2)$, respectively:
\begin{equation}
\vec{B}_3^\text{(q)}=A_3\left(-20
                              \frac{\red{Q}_3(r)}{r} S_1(\theta)\,,\
  -r^{-1}\oderiv{}{r}[rQ_3(r)]\oderiv{S_1(\theta)}{\theta}\,,\   0
                              \right)\,,
\end{equation}
where $A_3\approx0.133$, $S_1(\theta)=35\cos^{4}\theta-30\cos^2\theta+3$ and
\begin{equation}
\red{Q}_3(r)= \begin{cases}
    r^{-1/2}J_{9/2}(q_3r)\,,	&r \leq 1\,, \\
    r^{-5}J_{9/2}(q_3)\,,		&r > 1\,,
  \end{cases}
  \qquad q_3 \approx6.988\,,
\end{equation}
and
\begin{equation}
\vec{B}_4^\text{(q)}=A_4\left(0,\ 0,\ -\red{Q}_4(r)\oderiv{S_2(\theta)}{\theta}\right)\,,
\end{equation}
where $A_4\approx0.763$, $S_2(\theta)=5\cos^3\theta-3\cos\theta$ and
\begin{equation}
\red{Q}_4(r)=  \begin{cases}
    r^{-1/2}J_{7/2}(q_3r)\,,	&r \leq 1\,, \\
    r^{-4}J_{7/2}(q_3)\,, 		&r > 1\,.
  \end{cases} 
\end{equation}

\subsection{Anti-symmetric modes}\label{sec:free_anti}
The spherical components of magnetic field in a few leading anti-symmetric modes 
have the following form, illustrated in the bottom row of Fig.~\ref{fig:free_decay}. 

The mode that decays most slowly is poloidal, with $(n,l)=(1,1)$:
\begin{equation}
\vec{B}_1^\mathrm{(d)}=C_1\left(\frac{2}{r}Q_1(r)\,\cos\theta, \ 
-\frac{\sin\theta}{r}\oderiv{}{r}[rQ_1(r)], \  0 \right)\,,
\end{equation}
where  $C_1 \approx 0.346$ and
\begin{equation}
Q_{1}(r)=  \begin{cases}
    r^{-1/2}J_{3/2}(k_1r)\,, 	& r \leq 1\,, \\
    r^{-2}J_{3/2}(k_1)\,, 		& r > 1\,. 
  \end{cases}
  \qquad k_1=\pi\,.
\end{equation}

The next two modes $\vec{B}_2^\mathrm{(d)}$ and $\vec{B}_3^\mathrm{(d)}$,
poloidal and toroidal with $(n,l)=(3,1)$ and $(n,l)=(2,2)$,
respectively, form a degenerate pair:
\begin{align}
\vec{B}_2^\mathrm{(d)} &= C_2 \left(\frac{2\cos\theta}{r}
			(5\cos2\theta-1)Q_2(r)\right., \nonumber\\ 
	&\left.-\frac{\sin\theta}{r}(5\cos^2\theta-1)
    	\oderiv{}{r}\left[rQ_2(r)\right]\,, \ 0\right)\,,
\end{align}
where $C_2 \approx 0.250$ and
\begin{equation}
Q_2(r) =  \begin{cases}
    r^{-1/2}J_{7/2}(k_2r)\,, 	&r \leq 1\,, \\
    r^{-4}J_{7/2}(k_2)\,, 		&r > 1\,,
  \end{cases}
  \qquad k_2 \approx 5.763\,.
\end{equation}
The toroidal mode of the doublet has the form
\begin{equation}
\vec{B}_3^\mathrm{(d)}=C_3
				\left(0\,,\  0\,,\  Q_3(r)\,\sin\theta\,\cos\theta \right)\,,
\end{equation}
where $C_3 \approx 3.445$ and
\begin{equation}
Q_3(r) =  \begin{cases}
    r^{-1/2}J_{5/2}(k_2r)\,,	&r \leq 1\,, \\
    r^{-3}J_{5/2}(k_2r)\,,		&r > 1,. 
  \end{cases}
\end{equation}

The fourth antisymmetric mode is also poloidal, with ${(n,l)=(1,3)}$:
\begin{equation}
\vec{B}_4^\mathrm{(d)}=C_4\left(\frac{2}{r}Q_4(r)\,\cos\theta, \
\red{-}\frac{1}{r}\oderiv{}{r}[rQ_4(r)]\,\red{\sin\theta}, \  0 \right)\,,
\end{equation}
where  $C_4 \approx 0.244$\ and
\begin{equation}
Q_{4}(r)=  \begin{cases}
    r^{-1/2}J_{3/2}(k_4r)\,, 	& r \leq 1\,, \\
    r^{-2}J_{3/2}(k_4)\,, 		& r > 1\,, 
  \end{cases}
  \qquad k_4=2 \pi\,.
\end{equation}

\begin{figure}
 \centering
 \includegraphics{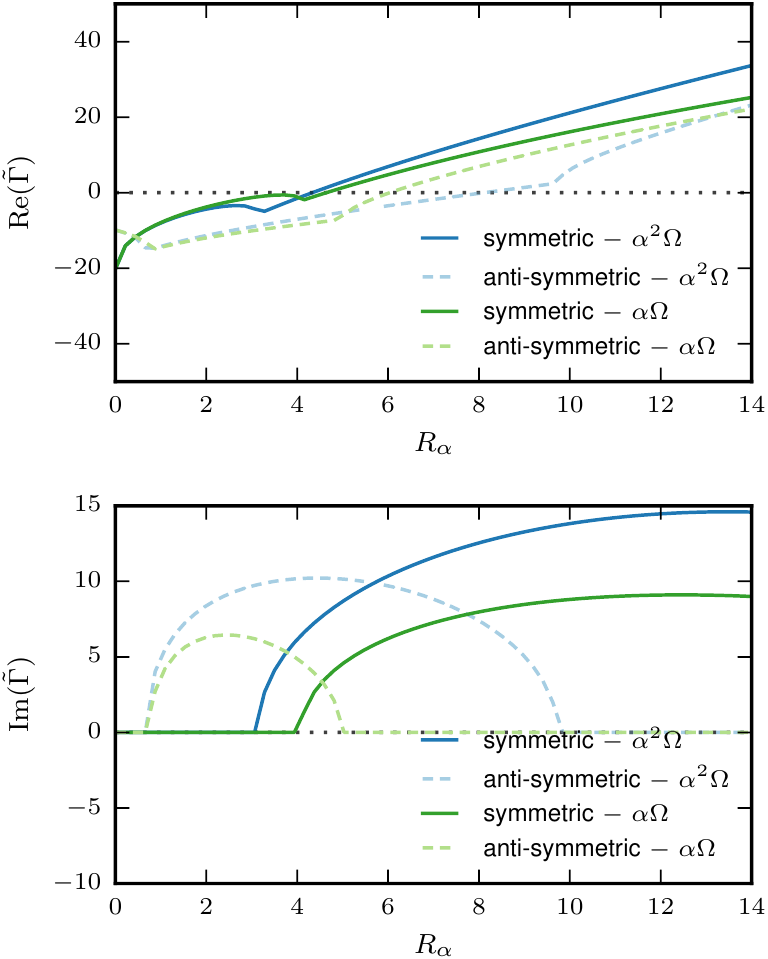}
 \caption{Growth rates $\re\Gamma$ and oscillation frequencies $\im\Gamma$  
 of the symmetric and anti-symmetric fastest-growing eigenmodes of the spherical 
 $\alpha\omega$-dynamo as a function of $\Rah$ with the remaining parameters
 fixed at their fiducial values. Dashed  curves show the results obtained for the
 $\alpha^2\omega$-dynamo, whereas solid curves are for the 
 $\alpha\omega$-dynamo. 
          }
 \label{fig:halo_growth_alphaOmega}
\end{figure}
\section{The $\alpha\omega$-dynamo approximation for the halo}\label{ap:alphaOmega}
In the main part of this paper, we use an $\alpha^2\omega$-dynamo model
for the halo, where the toroidal magnetic field is produced from the poloidal one
by both differential rotation and the $\alpha$-effect. The contribution of the
$\alpha^2$-dynamo to the generation of the toroidal field is usually weaker and 
therefore often neglected to simplify the solutions. To assess the consequences
of this approximation, here we provide solutions of the $\alpha\omega$-dynamo in 
the halo which should be compared with results presented in Section~\ref{MSH}.

For the $\alpha\omega$-dynamo, the operator defined in Eq.~\eqref{eq:What_alpha2omega}
is simplified to
\begin{equation}
    \What\vec{B}=\Rah\left(\nabla\times({\alpha}\vec{B})-[
    \nabla\times({\alpha}\vec{B})]_{\phi}\right)+\Roh\nabla\times(\vec{V}
    \times\vec{B})\,,
    \label{eq:What_alphaomega}
\end{equation}
thus  removing the contribution of the $\alpha$-effect to the azimuthal magnetic 
field. In Fig.~\ref{fig:halo_growth_alphaOmega}, similar to
Fig.~\ref{fig:halo_growth}, we show the resulting eigenvalues and expansion
coefficients of the perturbation solution. The difference between the two
solutions is noticeable.

The $\alpha^2$-dynamo is negligible in comparison with the $\alpha\omega$ 
mechanism when $|R_\omega/R_\alpha|\gg1$ or even $|R_\omega/R_\alpha^2|\gg1$ 
\citep{RuSoSh80}. For the fiducial value of parameters, 
$|R_\omega/R_\alpha|\simeq140$ in the disc and 50--100 in the halo 
(and $|\Roh/\Rah^2=10\text{--}50$).  
Thus, unlike the case of galactic discs 
with strong differential rotation, the $\alpha\omega$-approximation cannot be 
recommended for the halo because it neglects a potentially important part of the 
dynamo mechanism.

\section{Synchrotron emission and Faraday rotation}\label{ap:polarisation}
For the line of sight along the $x$-axis of a Cartesian reference frame
$\vec{r}=(x,y,z)$, the synchrotron emissivity at a wavelength $\lambda$ is 
derived as
\begin{equation}
 \epsilon(\vec{r},\lambda) \propto 
 	[B_y^2(\vec{r})+B_z^2(\vec{r})]^{(\kappa+1)/4} \lambda^{(\kappa-1)/2}\,,
\end{equation}
assuming a uniform distribution of cosmic ray electrons and the cosmic ray energy 
spectrum $N(E)\,\od E \propto E^{-\kappa}\, \od E$ with $\kappa=3$. The
Stokes parameters are computed as
\begin{align}
  I(y,z,\lambda) &= \int_{-\infty}^\infty \epsilon(x',y,z,\lambda)\,\dd x'\,,
  \label{eq:I}\\
  Q(y,z,\lambda) &=p_0 \int_{-\infty}^\infty \epsilon(x',y,z,\lambda)\, 
                \cos[2\psi(x',y,z)]\,\dd x'\,,
  \label{eq:Q}\\
  U(y,z,\lambda) &= p_0\int_{-\infty}^\infty \epsilon(x',y,z,\lambda)\, 
                \sin[2\psi(x',y,z)]\,\dd x'\,,
  \label{eq:U}
\end{align}
with the intrinsic polarisation degree $p_0=0.75$,
and the local polarisation angle $\psi(\vec{r})$ is obtained from
\begin{align}
 \psi(\vec{r}) =& \frac{\pi}{2}+\arctan\left[
 				\frac{B_z(\vec{r})}{B_y(\vec{r})}\right]\nonumber\\
      +& 0.81\rad \left(\frac{\lambda}{1\m}\right)^2\int_x^{\infty}
        \frac{n_\e(x',y,z)}{1\cm^{-3}}\,
        \frac{B_x(\bm{r}')}{1\muG}\,
        \frac{\dd x'}{1 \pc}\,,
\end{align}
where the thermal electron density $n_\e(\vec{r})$ is given by Eq.~\eqref{eq:ne}
in both the disc and the halo. The polarised intensity, observed 
polarisation angle and fractional polarisation follow as
\begin{equation}
 P = \sqrt{Q^2+U^2}\,,\quad \Psi = \tfrac{1}{2}\arctan(U/Q)\,, \qquad  p = P/I\,.
\end{equation}
The Faraday rotation measure is calculated as $\text{RM}=\partial\Psi/\partial(\lambda^2).$

\end{appendix}
\end{document}